\documentclass{aastex62}
\usepackage[utf8]{inputenc}
\usepackage{amsmath,amsfonts,amssymb}
\usepackage{subfigure}
\usepackage{multirow}
\usepackage{epstopdf}
\usepackage{graphicx,enumerate}
\usepackage{hyperref}
\usepackage{enumitem}
\usepackage{natbib}
\usepackage{xcolor}

\shorttitle{OVRO-LWA: Radio transient search}
\shortauthors{Anderson et al.}

\begin{document}
\title{New Limits on the Low-frequency Radio Transient Sky Using 31 hr of All-sky Data with the OVRO--LWA}

\author{Marin M. Anderson}
\affiliation{California Institute of Technology, 1200 E California Blvd MC 249-17, Pasadena, CA 91125, USA}

\author{Gregg Hallinan}
\affiliation{California Institute of Technology, 1200 E California Blvd MC 249-17, Pasadena, CA 91125, USA}

\author{Michael W. Eastwood}
\affiliation{California Institute of Technology, 1200 E California Blvd MC 249-17, Pasadena, CA 91125, USA}

\author{Ryan M. Monroe}
\affiliation{California Institute of Technology, 1200 E California Blvd MC 249-17, Pasadena, CA 91125, USA}

\author{Thomas A. Callister}
\affiliation{LIGO Laboratory, California Institute of Technology, Pasadena, CA 91125, USA}

\author{Jayce Dowell}
\affiliation{Department of Physics and Astronomy, University of New Mexico, Albuquerque, NM 87131, USA}

\author{Brian Hicks}
\affiliation{Naval Research Laboratory, Washington, DC 20375, USA}

\author{Yuping Huang}
\affiliation{California Institute of Technology, 1200 E California Blvd MC 249-17, Pasadena, CA 91125, USA}

\author{Namir E. Kassim}
\affiliation{Naval Research Laboratory, Washington, DC 20375, USA}

\author{Jonathon Kocz}
\affiliation{California Institute of Technology, 1200 E California Blvd MC 249-17, Pasadena, CA 91125, USA}

\author{T. Joseph W. Lazio}
\affiliation{Jet Propulsion Laboratory, California Institute of Technology, 4800 Oak Grove Dr, Pasadena, CA 91109, USA}

\author{Danny C. Price}
\affiliation{Department of Astronomy, University of California Berkeley, Berkeley CA 94720, USA}
\affiliation{Centre for Astrophysics \& Supercomputing, Swinburne University of Technology, PO Box 218, Hawthorn, VIC 3122, Australia}

\author{Frank K. Schinzel}
\affiliation{National Radio Astronomy Observatory, P.O. Box O, Socorro, NM 87801, USA}
\affiliation{Department of Physics and Astronomy, University of New Mexico, Albuquerque, NM 87131, USA}

\author{Greg B. Taylor}
\affiliation{Department of Physics and Astronomy, University of New Mexico, Albuquerque, NM 87131, USA}

\correspondingauthor{Marin M. Anderson}
\email{mmanders@astro.caltech.edu}

\begin{abstract}
We present the results of the first transient survey from the Owens Valley Radio Observatory Long Wavelength Array (OVRO--LWA) using 31\,hr of data, in which we place the most constraining limits on the instantaneous transient surface density at timescales of 13\,s to a few minutes and at frequencies below 100\,MHz. 
The OVRO--LWA is a dipole array that images the entire viewable hemisphere with 58\,MHz of bandwidth from 27 to 84\,MHz at 13\,s cadence. No transients are detected above a 6.5$\sigma$ flux density limit of 10.5\,Jy, implying an upper limit to the transient surface density of $2.5\times10^{-8}$\,deg$^{-2}$ at the shortest timescales probed, which is orders of magnitude deeper than has been achieved at sub-100\,MHz frequencies and comparable flux densities to date. The nondetection of transients in the OVRO--LWA survey, particularly at minutes-long timescales, allows us to place further constraints on the rate of the potential population of transients uncovered by \citealt{Stewart+2016}. From their transient rate, we expect a detection of $8.4^{+31.8}_{-8.0}$\,events, and the probability of our null detection is $1.9^{+644}_{-1.9}\times10^{-3}$, ruling out a transient rate $>1.4\times10^{-4}\,\text{days}^{-1}\,\text{deg}^{-2}$ with 95\% confidence at a flux density limit of 18.1\,Jy, under the assumption of a flat spectrum and wide bandwidth. We discuss the implications of our nondetection for this population and further constraints that can be made on the source spectral index, intrinsic emission bandwidth, and resulting luminosity distribution.
\end{abstract}

\section{Introduction}\label{introduction}
Exploration of the time domain plays a critical role in the field of astronomy, as a window into the dynamic, often cataclysmic or explosive, processes that mark a highly nonstatic universe. Time domain radio astronomy, in particular, serves as a unique probe of astrophysical processes that can be otherwise inaccessible at other wavelengths, due either to extrinsic factors, such as dust obscuration, or intrinsic factors, such as emission that is beamed or unique to radio frequencies.

Recent advances in the survey speed and sensitivity of radio interferometers has facilitated the expansion of radio transient science from the realm of follow-up of transient detections at shorter wavelengths, such as extragalactic, incoherent synchrotron events such as gamma-ray bursts (GRBs), supernovae, and tidal disruption events~\citep[e.g.,][]{Soderberg+2010, Zauderer+2011, Chandra+Frail2012}, toward wide-field, nontargeted synoptic imaging surveys exploring a wide range of phase space in the dynamic radio sky~\citep[e.g.,][]{Mooley+2016}. Incoherent synchrotron-powered transients, due to both the lower luminosities from self-absorption and typically slow timescale of evolution at low frequencies, are better recovered at GHz frequencies~\citep[see][]{Metzger+2015}. However, surveys at low ($<350$\,MHz) frequencies benefit from the many sources of coherent emission that fill radio transient phase space~\citep[see][]{Cordes+2004, Macquart+2015}. Coherent emission is extremely bright, and can be variable on timescales as short as nanoseconds~\citep{Hankins+2003}. In addition, many of the source populations that have been identified to date either emit exclusively at low frequencies, e.g., cyclotron maser-powered auroral radio emission from the magnetized planets in the solar system~\citep{Zarka1998}, or exhibit emission with steep negative spectral indices~\citep{Lorimer+1995,Kramer+1999} that motivate surveys conducted at lower frequencies. Surveys at low frequencies also benefit from the inherently wide fields of view and fast survey speeds of dipole arrays. This is increasingly the case with the rise of low frequency telescopes such as the Long Wavelength Array~\citep[LWA;][]{Ellingson+2009}, the Murchison Widefield Array~\citep{Tingay+2013}, the Low-Frequency Array~\citep[LOFAR;][]{vanHaarlem+2013}, and now the Owens Valley Radio Observatory Long Wavelength Array (OVRO--LWA). This also includes the upgrade of low-frequency receivers on existing dish arrays, such as the Very Large Array (VLA) Low Band Ionospheric and Transient Experiment~\citep[VLITE;][]{Polisensky+2016} and the Giant Metrewave Radio Telescope~\citep[GMRT;][]{Intema+2017}.

At extragalactic distances, potential transient sources at low frequencies include bright, coherent pulses predicted to accompany GRBs and neutron star mergers~\citep[see][and references therein]{Anderson+2018}, as well as fast radio bursts (FRBs), highly dispersed millisecond pulses that have been detected at frequencies as low as 400\,MHz~\citep{Amiri+2019}. The degree of scatter broadening from the current sample of FRBs indicates the potential for detecting this population even at very low frequencies~\citep{Ravi2019}. At galactic distances, potential transient sources at low frequencies include stellar flares and coherent radio bursts~\citep{Spangler+Moffett1976, Bastian1990, Lynch+2017, Villadsen+2018}, for which our Sun serves as a prototype~\citep{Bastian+1998}, as well as radio aurorae on brown dwarfs~\citep{Hallinan+2015, Kao+2016}, a phenomenon that likely also extends down in mass to exoplanets~\citep{Lazio+2004, Kao+2018}, and for which the magnetized planets in our solar system serve as a prototype~\citep{Zarka1998}. Within Earth's atmosphere, potential transient sources include radio emission from meteor afterglows recently discovered at low frequencies~\citep{Obenberger+2014}, as well as radio emission from cosmic-ray showers and neutrinos~\citep{Falcke+Gorham2003}.

Transient surveys carried out by the aforementioned low-frequency facilities continue to place more sensitive and constraining limits on the radio transient sky below 350\,MHz, across a wide range of timescales. Table~\ref{tab:previoussurveys} summarizes the parameters and results of previous blind, nontargeted transient surveys at low frequencies ($<350\,\text{MHz}$). From these surveys, there have been seven distinct transient discoveries, all of which remain mysterious in origin, with no identified progenitors and no corresponding larger population yet uncovered.
\begin{enumerate}
\item Three transient sources were discovered during separate monitoring campaigns of the galactic center, thus the label Galactic Center Radio Transient (GCRT). The first of these, GCRT\,J1746-2757, found in 330\,MHz VLA observations of the galactic center, evolved on the timescale of a few months and reached $\sim200\,\text{mJy}$ flux densities~\citep{Hyman+2002}. GCRT\,J1745-3009, also found at 330\,MHz with the VLA and subsequently redetected with the GMRT, was a coherent, steep-spectrum source, exhibiting $\sim1\,\text{Jy}$ bursts of approximately 10\,minutes in duration, that were detected repeatedly with a 77\,minute period~\citep{Hyman+2005, Hyman+2007}. Finally, GCRT\,J1742-3001 was discovered at 235\,MHz with the GMRT. It reached $\sim100\,\text{mJy}$ flux densities, and evolved on a timescale of approximately a few months~\citep{Hyman+2009}.
\item \citealt{Jaeger+2012} detected the source J103916.2+585124 in 325\,MHz VLA archival observations that were selected based on their coverage of the \textit{Spitzer-Space-Telescope} Wide-area Infrared Extragalactic Survey (SWIRE) Deep Field. The source reached 2\,mJy flux densities with a duration of $\sim12\,\text{hr}$, and has no identified progenitor, despite extensive multiwavelength coverage of the field in which it was detected.
\item TGSSADR\,J183304.4--384046 was found in comparison of the 150\,MHz GaLactic and Extragalactic All-sky MWA (GLEAM) and TIFR GMRT Sky Survey (TGSS) catalogs. Given the 180\,mJy flux density and long (1--3\,yr) timescale of evolution of this source, this candidate transient may be an example of an extragalactic synchrotron transient detected at low frequencies, the first of many such transients that may be uncovered by the Square Kilometre Array~\citep[see][]{Metzger+2015} -- follow-up observations are ongoing~\citep{Murphy+2017}.
\item From 10,240\,hr of simultaneous monitoring with LWA1 and LWA Sevilleta~\citep[LWA--SV;][]{Cranmer+2017}, a possible transient was discovered at 34\,MHz with an approximately 830\,Jy flux density and a duration of 15--20\,s~\citep{Varghese+2019}, based on its simultaneous detection by both LWA stations.
\item The transient detection that is most relevant to this work was detected by \citealt{Stewart+2016} in LOFAR Multifrequency Snapshot Sky Survey (MSSS) data. The source, ILT\,J225347+862146, was detected at 60\,MHz and reached peak flux densities of 15--25\,Jy over a timescale of 11\,minutes. Although the origin of this transient emission remains a mystery, with no identified higher-energy counterpart, the flux density and implied rate of this event suggest that this may be a dominant population in the low-frequency transient sky, and one to which the OVRO--LWA is sensitive.
\end{enumerate}

\begin{deluxetable*}{ccCCCcccC}[htb!]
	\tabletypesize{\scriptsize}
	\tablecolumns{9}
	\tablecaption{Radio transient surveys at low frequencies ($<350$\,MHz).\label{tab:previoussurveys}}
	\tablehead{ 
	(1)				& (2)					& (3)					& (4)					& (5)					& (6)							& (7)					& (8)							& (9)\\
	\colhead{Reference} 	& \colhead{Instrument} 	& \colhead{Center Frequency} 	& \colhead{Bandwidth}	& \colhead{Search Sensitivity} 	& \colhead{Timescale Searched} 	& \colhead{$\Omega_\text{FOV}$} 	& \colhead{$N_{\text{epochs}}$}	& \colhead{$N_{\text{transients}}$}\\
					&  					& \colhead{(MHz)} 		& \colhead{(MHz)}		& \colhead{(Jy)} 		& 							& \colhead{(deg$^{2}$)} 	&  					 		& }

	\startdata
	0	& LWA1 \& LWA--SV	& 34, 38	& 0.1	& 246	& 5\,s, 15\,s, 60\,s	& 10,300	& $7.4\times10^6$	& 1 \\
	&&&& \vspace{-0.2cm} 10.5, 13.8, 18.1, & 13\,s, 39\,s, 2\,minutes, && 8586, 2862, 954,  & \\
	\vspace{-0.2cm}1	& OVRO--LWA	& 56	& 58	& & & 17,045 & & 0 \\
	&&&& 20.8, 19.3, 19.7 & 6\,minutes, $<1$\,day, $<35$\,days	&& 318, 1, 1 & \\
	&&&& \vspace{-0.2cm} 36.1, 21.1, 7.9, & 30\,s, 2\,minutes, 11\,minutes, && 41,340, 9262, 1897 &  \\
	\vspace{-0.2cm}2	& LOFAR	& 60	& 0.195 & & & 176.7	& & 1 \\
	&&&& 5.5, 2.5 & 55\,minutes, 297\,minutes && 328, 32 &  \\
	&& \vspace{-0.2cm} 38, 52 && 540, 230 &&& $6.1\times10^6$, $1.4\times10^6$ &  \\
	\vspace{-0.2cm}3	& LWA1	&& 0.075	&& 5\,s & 8353.7	&& 0 \\
	&& 74 && 570 &&& $1.0\times10^6$ &  \\
	4	& LWDA			& 73.8				& 1.6					& 2~500				& 300\,s						& 10,000				& 29,437						& 0 		\\
	5	& LOFAR			& 149				& 0.781				& 0.5					& 11\,minutes						& 11.35				& 26							& 0 		\\
	6	& LOFAR			& 150				& 48					& 0.3					& 15\,minutes, 100\,days					& 15.48				& 151						& 0		\\
	7	& MWA-32T		& 154				& 30.72				& 5.5					& 26\,minutes, 1\,yr					& 1430				& 51							& 0		\\
	8	& MWA			& 150				& \nodata				& 0.1					& 1-3\,yr						& 16,230				& 2							& 1		\\
	&&&&& \vspace{-0.2cm} 28\,s, 5\,minutes, 10\,minutes, 1\,hr, 2\,hr, &&& \\
	\vspace{-0.2cm}9	& MWA	& 182			& 7.68				& 0.285				& 							& 452				& 10,122						& 0		\\
	&&&&& 1\,day, 3\,days, 10\,days, 30\,days, 90\,days, 1\,yr &&& \\
	10	& MWA			& 182				& 30.72				& 0.02, 0.200			& 1\,hr, 30\,days					& 186				& 652, 28						& 0		\\
	11	& GMRT			& 235				& 15					& 0.003 - 0.01			& 1\,day -- 2\,yr					& 3.2					& 20							& 1		\\
	12	& VLA			& 325				& 12.5				& 0.0021				& 12\,hr						& 6.5					& 6							& 1		\\
	13	& VLA			& 330				& 3.1					& 64.8				& $\sim$ 10\,yr					& 4.9					& 2							& 1		\\
	14	& VLA			& 330				& 6.2					& 0.05				& 5\,minutes						& 7					& 440			& 1		\\
	15	& VLITE			& 340				& 64					& 0.1					& 10\,minutes -- 6\,hr					& 5.5					& 2799						& 0		\\
	\enddata
\tablerefs{(0) \citealt{Varghese+2019}; (1) This work; (2) \citealt{Stewart+2016}; (3) \citealt{Obenberger+2015}; (4) \citealt{Lazio+2010}; (5) \citealt{Cendes+2014}; (6) \citealt{Carbone+2016}; (7) \citealt{Bell+2014}; (8) \citealt{Murphy+2017}; (9) \citealt{Rowlinson+2016}; (10) \citealt{Feng+2017}; (11) \citealt{Hyman+2009}; (12) \citealt{Jaeger+2012}; (13) \citealt{Hyman+2002}; (14) \citealt{Hyman+2005}; (15) \citealt{Polisensky+2016}.}
\tablecomments{Summary of transient surveys at low frequencies ($<350$\,MHz) conducted to date, ordered from lowest to highest frequency. Note that only nontargeted, ``blind" surveys are included here. In column 5, the search sensitivity refers to the detection threshold that was specified by each survey for transient detection -- when not specified, a $6\sigma$ threshold is assumed. When only one value is given for $N_\text{epochs}$ for a survey probing a range of timescales, the value applies to the shortest timescale reported. For \citealt{Hyman+2002}, we consider only the two epochs that were involved in the initial discovery of GCRT\,J1746-2757, and exclude any follow-up analysis that targeted the field post-discovery.}
\end{deluxetable*}

In order to elucidate the nature of the potential population of sources revealed by the transient of \citealt{Stewart+2016}, as well as to explore the low-frequency radio sky for other transient phenomena, we have conducted a survey with the OVRO--LWA from 27 to 84\,MHz, imaging the entire viewable sky ($\sim1.6\pi\,\text{sr}$) every 13\,s with approximately 10$'$ spatial resolution, to search for transient sources on timescales ranging from 13\,s to $\sim$days, using 31\,hr of observations. The survey observations and description of the OVRO--LWA are given in Section~\ref{observations}. In Section~\ref{analysis}, we describe our transient detection pipeline, with the results of the survey detailed in Section~\ref{results}. The upper limits on the transient surface density placed by the nondetection of transients in our survey are given in Section~\ref{discussion}, as well as the implications of our nondetection on the potential population of transients similar to that detected by \citealt{Stewart+2016}. We conclude in Section~\ref{conclusion}, where we also outline the future of OVRO--LWA transient science.

\section{Observations}\label{observations}

\subsection{OVRO--LWA}\label{ovro-lwa}
The OVRO--LWA is a low-frequency, dual-polarization dipole array currently under development at the Owens Valley Radio Observatory (OVRO) in Owens Valley, California, operating between 27--84\,MHz. The completed array will consist of 352 elements -- 251 elements contained within a 200\,m diameter compact core, and 101 elements spread across maximum baselines of 2.5\,km. The OVRO--LWA operates by cross-correlating the signals from all elements, providing a snapshot imaging capability with the full-sky field-of-view (FOV) of a dipole antenna, roughly 5$'$ spatial resolution at a cadence of a few seconds, with 100\,mJy sensitivity across an instantaneous bandwidth of approximately 60\,MHz.

At the time of the observations described here, the OVRO--LWA was in ``stage II" of development, incorporating the 251-element core and a 32-element Long Baseline Demonstrator Array (LBDA) spread across maximum baselines of 1.5\,km. The Large-Aperture Experiment to Detect the Dark Age (LEDA) correlator~\citep{Kocz+2015} takes 512 inputs (256 antennas $\times$ 2 polarizations) and correlates all signals across a 27-84\,MHz band (2400 channels at 24\,kHz resolution) with a 13\,s cadence. The stage II OVRO--LWA began operations in 2016 December, and early science includes: constraints on the sky-averaged H\,I absorption trough at $16<z<34$~\citep{Price+2018}; high-fidelity, low-frequency maps for use in foreground filtering techniques, as well as the first constraints on the angular power spectrum of redshifted H\,I at low frequencies~\citep{Eastwood+2018, Eastwood+2019}; the first detection of high-energy cosmic rays without reliance on triggers from particle detectors~\citep{Monroe+2019}; and constraints on coherent emission mechanisms associated with short gamma ray bursts~\citep[GRBs;][]{Anderson+2018}.

\subsection{The 31\,hr Data Set}\label{thesurveydata}
The first dedicated survey observation with the stage II OVRO--LWA was conducted on 2017 February 17, for 28\,hr of continuous observations from 12:00 UTC through 2017 February 18 16:00 UTC. Data were recorded at 13\,second integrations, with 2398 frequency channels spanning 27.38 -- 84.92\,MHz (24\,kHz frequency resolution). The cadence was chosen because a sidereal period is almost exactly an integer multiple of 13\,s, and this makes possible the subtraction of sidereally separated integrations as a means of searching for transients with sensitivity to all timescales up to a sidereal day. The OVRO--LWA is a zenith-pointing telescope, with each snapshot image covering the full visible hemisphere -- the 28\,hr run consists of 7756 contiguous full sky snapshots. In addition to this data set, a 3\,hr observation (832 contiguous full sky snapshots) from 2017 January 12 01:02 to 04:02 UTC that was conducted as part of a short GRB follow-up program~\citep{Anderson+2018} was included (with a configuration nearly identical to that of the 28\,hr data set), to bring the amount of data used in the transient search to 31\,hr (see Table~\ref{tab:dataset}).

\begin{deluxetable}{cccc}[htb!]
	\tabletypesize{\scriptsize}
	\tablecolumns{4}
	\tablecaption{The Observations and Corresponding Calibration Times that Comprise the 31 hr Data Set.\label{tab:dataset}}
	\tablehead{
		(1)					& (2)						& (3)							& (4)						\\
		\colhead{Start Time (UTC)}	& \colhead{End Time (UTC)}	& \colhead{Number of Snapshots}	& \colhead{Calibration (UTC)} 	\\
			}
	\startdata
		2017 Jan 12 01:02:02		& 2017 Jan 12 04:02:05		& 832						& 2017 Jan 11 20:26:39		\\
		2017 Feb 17 11:59:54		& 2017 Feb 18 16:00:09		& 7756						& 2017 Feb 17 18:02:10		\\
	\enddata
\end{deluxetable}

\subsection{Calibration and Imaging}\label{calibrationandimaging}
The raw visibility data sets produced by the LEDA correlator are converted into the standard Common Astronomy Software Applications~\citep[CASA;][]{McMullin+2007} measurement set table format~\citep{vanDiepen2015}. Prior to calibration, antenna autocorrelation spectra are inspected and antennas are flagged if the median spectral power of an antenna is not within two standard deviations of the median power across all antennas. This is sufficient for identifying all antennas with issues along the signal path that resulted in a loss of, or sufficient distortion of, sky signal. For this data set, 50 out of 256 antennas were flagged. An additional set of flags are applied in baseline space, to avoid the effects of cross-coupling in adjacent lines in the analog signal path, and to remove a subset of bad baselines that are manually selected from the visibilities, resulting in the removal of less than 2\% of baselines. Channel flags are generated on a per integration basis, by identifying channels with visibility amplitudes that are outliers in their mean and max amplitudes relative to the rest of the band. This typically results in fewer than 240 out of 2398 channels flagged. In total, approximately 40\% of visibilities are flagged per integration, the majority being due to the large number of antenna flags. However, because the array is still dominated by confusion noise and other systematics even at the shortest timescales, the removal of such a large fraction of visibilities does not impact the sensitivity of the array.

The data are calibrated using a simplified sky model consisting of the two brightest sources in the sub-100\,MHz sky: the radio galaxy Cygnus (Cyg) A and the supernova remnant Cassiopeia (Cas) A, using the model flux and spectral indices given in~\citealt{Baars+1977} for the former and~\citealt{Perley+Butler2017} for the latter, extrapolated down to 28 MHz. The complex (amplitude and phase) antenna gains are determined on a per channel basis from the two-source Cyg A-Cas A sky model using the CASA \texttt{bandpass} task, using baselines greater than 15 wavelengths in order to mitigate the effect of diffuse galactic synchrotron emission. The calibration solutions are derived from a single integration, and they remain sufficiently stable for roughly 24\,hr. Thus, only one set of calibration solutions are generated per observation (see Table~\ref{tab:dataset}), from an integration when Cyg A is at its highest elevation in the beam ($\sim87^\circ$). We note that the two-source model used in the calibration process constitutes a grossly incomplete sky model, but it does provide sufficient signal-to-noise (S/N) to solve for a set of complex gains. Future surveys will incorporate a more complete sky model (generated from the full-sky maps of~\citealt{Eastwood+2018}), with calibration solutions derived from visibilities averaged over multiple integrations.

\begin{figure*}[htb!]
\begin{center}
	\subfigure[]{\includegraphics[width=0.7\textwidth]{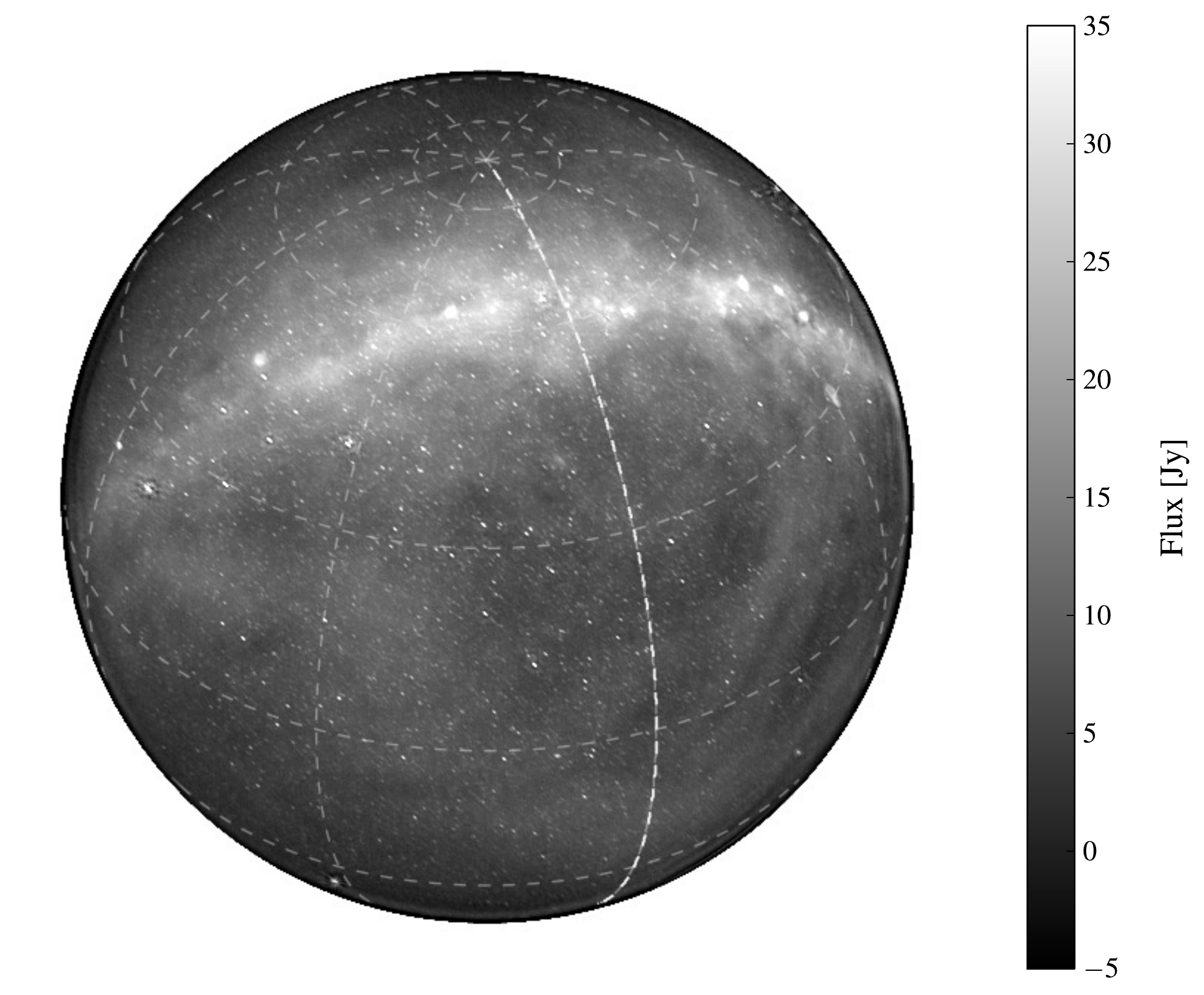}}
	\subfigure[\label{fig:fullskybeam}]{\includegraphics[width=0.7\textwidth]{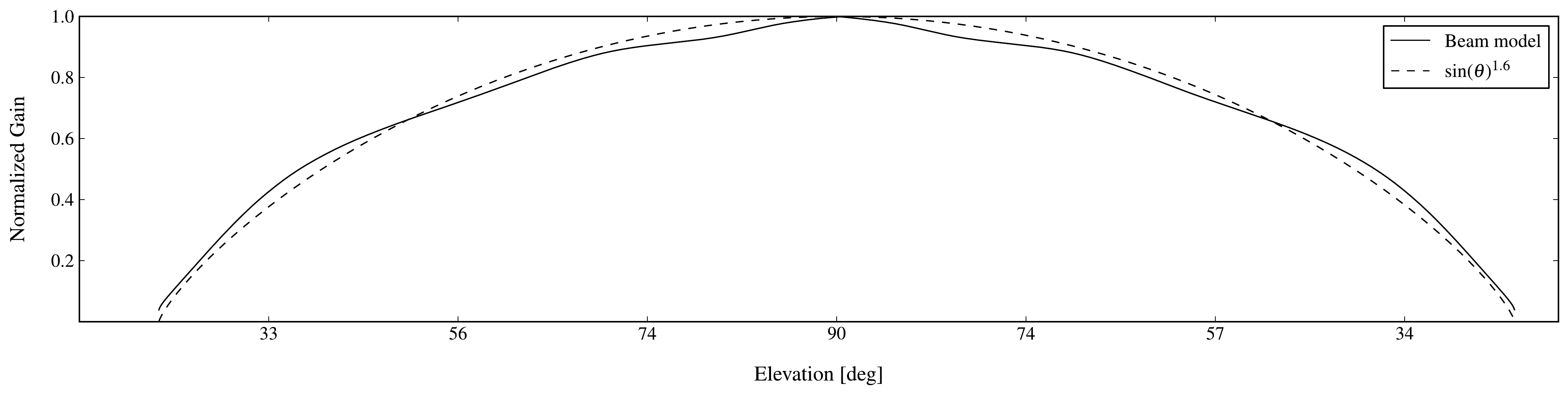}}
	\caption{Snapshot full-sky image from the OVRO--LWA at 2017 January 12 02:02:05 UTC, with dashed contours denoting lines of constant R.A.~and decl.~(a), and a cut through the primary beam at an azimuth of $0^\circ$, comparing a simulated LWA dipole beam with ground screen to the standard dipole approximation given by $\sin^{1.6}(\theta_\text{el})$, as a function of elevation angle (b). The beam model is symmetric, such that an orthogonal cut through the beam at an azimuth of $90^\circ$ would yield an identical normalized gain pattern. The entire field-of-view down to an elevation angle of 10$^\circ$ is searched in the transient pipeline. In the snapshot image, zenith is located in the center, with the horizon represented by the perimeter of the circle. The diffuse emission is galactic synchrotron emission. In a single snapshot image, more than 2000 point sources are detected above the local $5\sigma$ threshold (see \S\ref{analysis}).}
	\label{fig:fullsky}
\end{center}
\end{figure*}

Additional direction-dependent calibration and source subtraction (peeling) toward Cyg A and Cas A are performed on single-integration timescales (when either of the two sources is above the horizon), using visibilities with baselines greater than 10 wavelengths to derive the direction-dependent calibration. This is necessitated by variations in the antenna gain pattern between individual dipole beams. These variations are caused by mutual coupling between adjacent antennas, and to a lesser degree, by deviations in antenna orientation and ionospheric fluctuations. Peeling solutions are also derived on a per integration basis for a generic source in the near-field of the array. This is done to remove the effects of a stationary noise pattern in the data, which has distinct spectral, temporal, and spatial characteristics, likely caused by a combination of common-mode pickup and cross-talk in the analog electronics. Peeling is performed using the \texttt{TTCal} calibration software package developed for the OVRO--LWA~\citep{eastwood_michael_w_2016_1049160}. Imaging and deconvolution are done with \texttt{WSClean}~\citep{Offringa+2014}. The full FOV is imaged over 4096 $\times$ 4096 pixels, with a pixel scale of 1.$'$875 and using a robust visibility weighting of 0~\citep{Briggs1995}. At the time of these observations, a set of northeast LBDA antennas were not operational, resulting in an abnormally elongated synthesized beam with a major axis of 29$'$, a minor axis of 13.$'$5, and a position angle of 50$^\circ$. Figure \ref{fig:fullsky} shows an example 13\,s snapshot full sky image taken from the 31\,hr data set, using the full 58\,MHz of bandwidth, and an approximation of the primary beam as a function of elevation angle.

The transient detection by \citealt{Stewart+2016} in 400\,hr of LOFAR MSSS data with 195\,kHz of bandwidth at 60\,MHz, in a field centered on the north celestial pole, represents the detection of a potentially new and exciting transient source population. From this event, \citealt{Stewart+2016} report a transient rate of $3.9^{+14.7}_{-3.7}\times10^{-4}\,\text{d}^{-1}\,\text{deg}^{-2}$ at a flux density level of 7.9\,Jy. This corresponds to an expected detection of $8.4^{+31.8}_{-8.0}$ events in the 31\,hr OVRO--LWA transient survey, under the assumption that the transient emission is broadband in nature. Investigating and verifying this population therefore represents a primary goal of the OVRO--LWA transient survey. One of the timescales probed using the full 31\,hr of the OVRO--LWA transient data set is six minutes, which is able to capture the relevant timescale over which the \citealt{Stewart+2016} transient showed an appreciable increase in flux (the total duration of the event was 11 minutes, but it showed an approximately 100\% increase in flux over a two-minute period).

\section{The Transient Pipeline}\label{analysis}
\subsection{Timescales}\label{timescales}
The full 31\,hr data set comprising the OVRO--LWA transient survey was searched across a range of timescales by performing subtractions in image space of either sequential integrations or a sliding boxcar of width determined by the number of integrations in a given timescale, in order to remove the diffuse galactic emission and nonintrinsically varying sources and to generate ``difference" images that are sensitive to sources which are varying on the timescale given by the integration spacing. Sequential image subtraction was used to probe timescales of 13\,s (single integrations), 39\,s (3 integrations), 2\,minutes (9 integrations), and 6\,minutes (27 integrations). In addition, sidereal subtraction using 3\,hr of data with overlapping sidereal coverage between the 2017 January 12 and 2017 February 17 observations, as well as 4\,hr of data with overlapping sidereal coverage within the 2017 February 17 observations, was used to search for transients on all timescales simultaneously (up to 36\,days and 24\,hr, respectively).

\subsection{Sensitivity}\label{sensitivity}
The difference images are generated by selecting snapshots separated by $N$ integrations, matching the phase centers between the two snapshots to partially account for sky rotation, and subtracting the resulting images. Images are not deconvolved prior to subtraction, due to computational limitations associated with deconvolution of many tens of thousands of all-sky images. In addition, because we are not averaging visibilities in time when performing the transient search on longer timescales, and instead selecting two snapshot images separated by a given timescale, there is no gain in sensitivity (by a factor $\propto\sqrt{t_\text{int}}$) for difference images probing longer timescales. The average noise actually increases toward longer timescales, largely due to limitations associated with bright source sidelobes and incomplete sky subtraction between widely spaced integrations. The direction dependent calibration and subtraction (peeling) that was described in Section~\ref{calibrationandimaging} above is performed on the two brightest sources in the low-frequency sky (Cas A and Cyg A). The effects that necessitate the peeling of these sources are also present for the remaining sources in the FOV, but computational limitations and S/N requirements for the peeling process prevent the removal of the next set of brightest sources in the low-frequency sky (e.g., Taurus A and Virgo A). The result is sidelobe artifacts that cannot be successfully peeled out. These create a noise floor that is above the expected thermal noise on single integration timescales, and is therefore the limiting factor when averaging visibilities on longer timescales. The direction-dependent effects inherent to this issue are almost entirely dominated by beam-to-beam variation between dipoles, and the eventual mapping of the individual OVRO--LWA dipole beams will allow significant improvement in noise for difference image searches on longer timescales.

\begin{figure*}[htb!]
\begin{center}
	\includegraphics[width=0.5\textwidth]{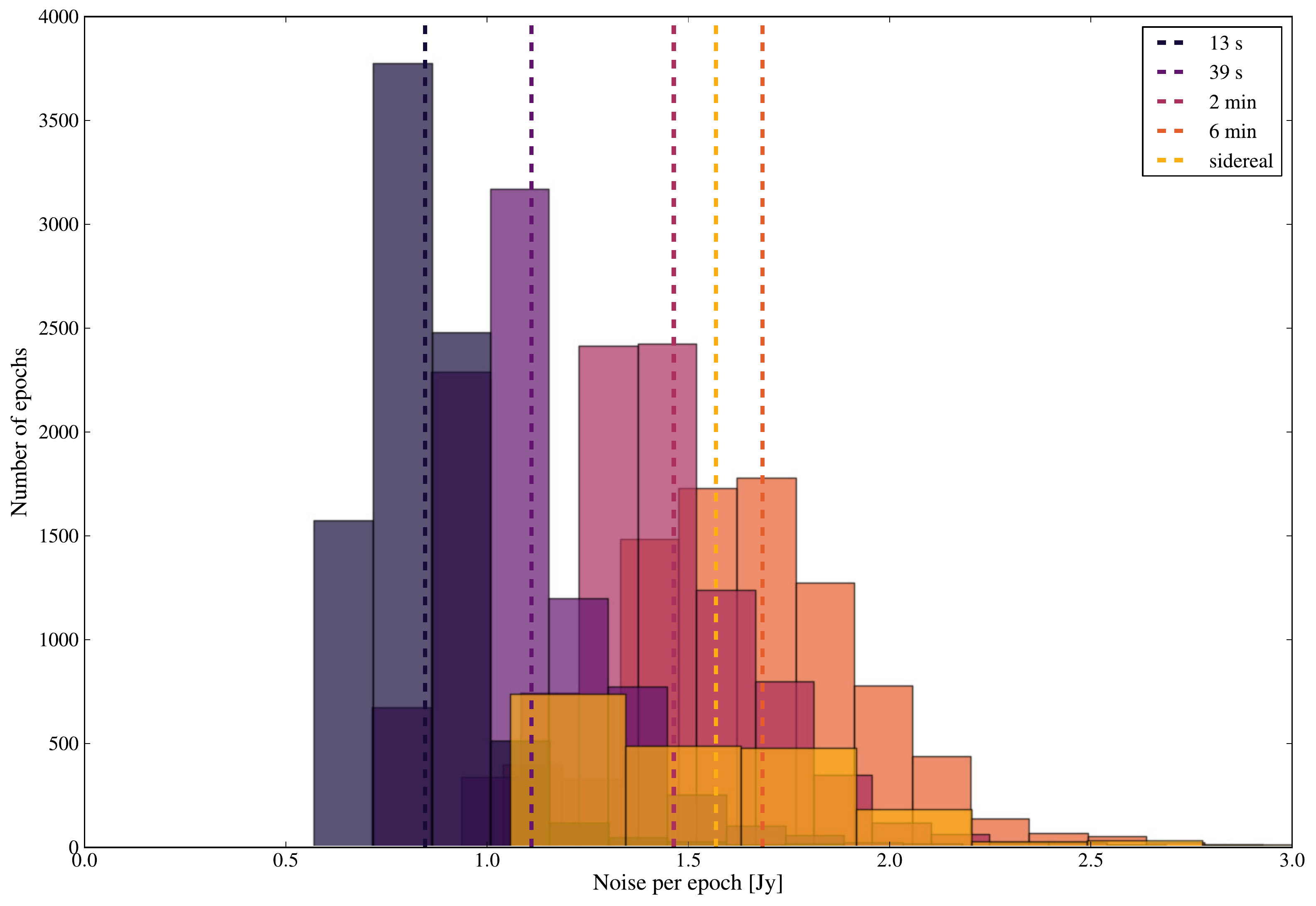}
	\caption{Histogram of the noise, as measured from the pixels in the $15^\circ$ region surrounding zenith, in each difference image epoch, for all timescales probed. The dashed lines show the mean value of the noise for each timescale.}
	\label{fig:noisehistogram}
\end{center}
\end{figure*}

The other factor that must be taken into account when computing the sensitivity of our survey is the fact that our transient pipeline searches over nearly the entire primary beam FOV, down to an elevation angle of $10^\circ$. The result is nonuniform sensitivity over our $\sim$17,000$\,\text{deg}^2$ FOV. To account for this, the noise is measured in the central regions of each difference image surrounding zenith. The effect of the primary beam is approximated as 1/sin$^{1.6}\theta_\text{elevation}$~(see Figure~\ref{fig:fullskybeam}), and the noise we measure across the whole field for each epoch of the survey is approximated as the noise at zenith times the mean of the primary beam pattern weighted by the fractional area of the FOV as a function of elevation angle. Taking into account the $6.5\sigma$ threshold of the transient search pipeline, this results in sensitivities at each timescale as reported in Table~\ref{tab:limits}. Because we are using a single measurement of the noise at zenith to approximate our sensitivity across the entire FOV, we are necessarily neglecting subtle variations in the noise across the image that are due to things like bright source sidelobes, etc. However, given the large number of epochs over which we are computing our survey sensitivity, these variations are not likely to significantly alter our noise values.

Table~\ref{tab:noise} gives the average RMS noise and number of epochs for each timescale searched, and Figure~\ref{fig:noisehistogram} shows the spread in noise for each timescale. We note that the noise in a difference image is equal to the quadrature sum of the noise for the two images from which it was generated (i.e., $\sigma_\text{diff} = \sqrt{\sigma_t^2 + \sigma_{t+\Delta t}^2}$). However, this is still an improvement in noise over the confusion-limited single snapshot images~\citep{Cohen2004}.

\begin{deluxetable}{cccc}[htb!]
	\tabletypesize{\scriptsize}
	\tablecolumns{3}
	\tablecaption{The Average Noise in the Sequentially Subtracted Images for Each of the Transient Timescales Searched.\label{tab:noise}}
	\tablehead{
		(1)					& (2)									& (3)								\\
		\colhead{Timescale}		& \colhead{Average RMS (Jy beam$^{-1}$)}	& \colhead{Number of Epochs}		 	\\
			}
	\startdata
		13~s					& 0.85								& 8586							\\
		39~s					& 1.11								& 8582							\\
		2~minutes				& 1.56								& 8570							\\
		6~minutes				& 1.68								& 8534							\\
		sidereal				& 1.57								& 1960							\\
	\enddata
\tablecomments{The noise is measured from a region centered on zenith with a radius of approximately $15^\circ$. Single integration (13\,s) images are consistent with thermal noise. There is an increase in noise on longer timescales, due to the fact that the difference images are generated by subtracting non-deconvolved images generated from visibilities at different local sidereal times (LSTs), rather than averaging in time (see Section~\ref{sensitivity}).}
\end{deluxetable}

\subsection{Source Extraction}\label{sourceextraction}
\begin{figure*}[htb!]
\begin{center}
	\subfigure[]{\includegraphics[width=0.49\textwidth]{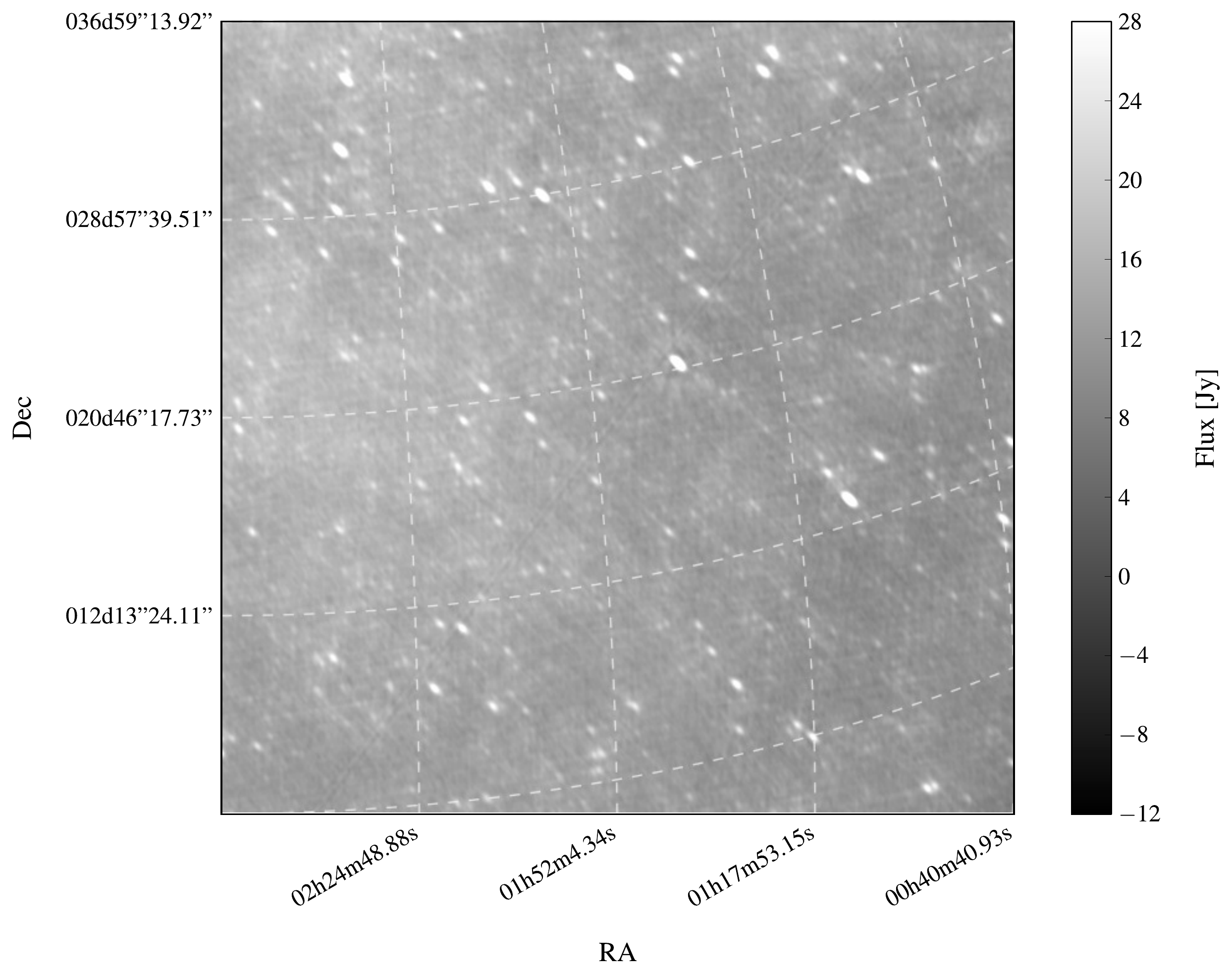}}
	\subfigure[]{\includegraphics[width=0.49\textwidth]{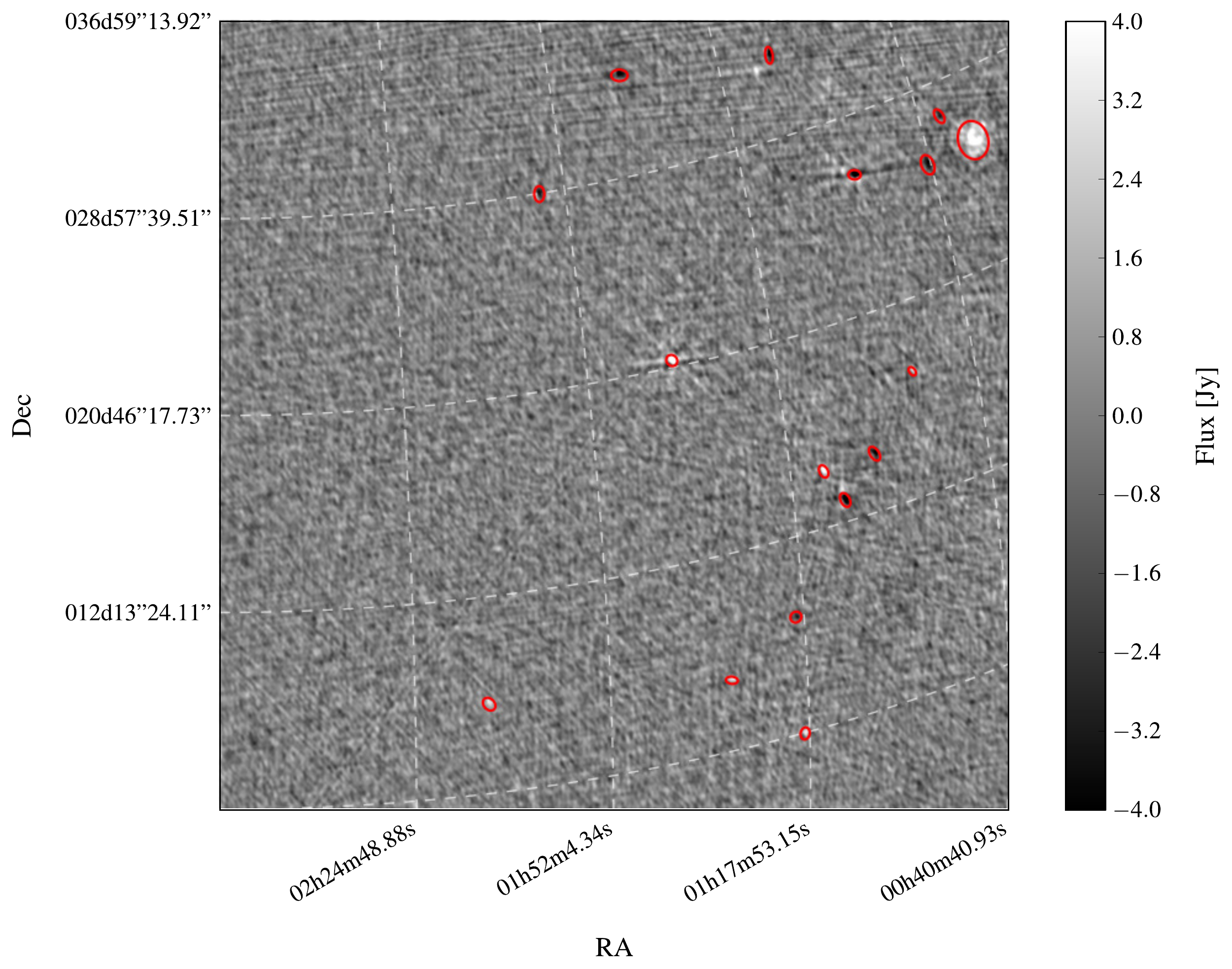}}
	\subfigure[]{\includegraphics[width=1\textwidth]{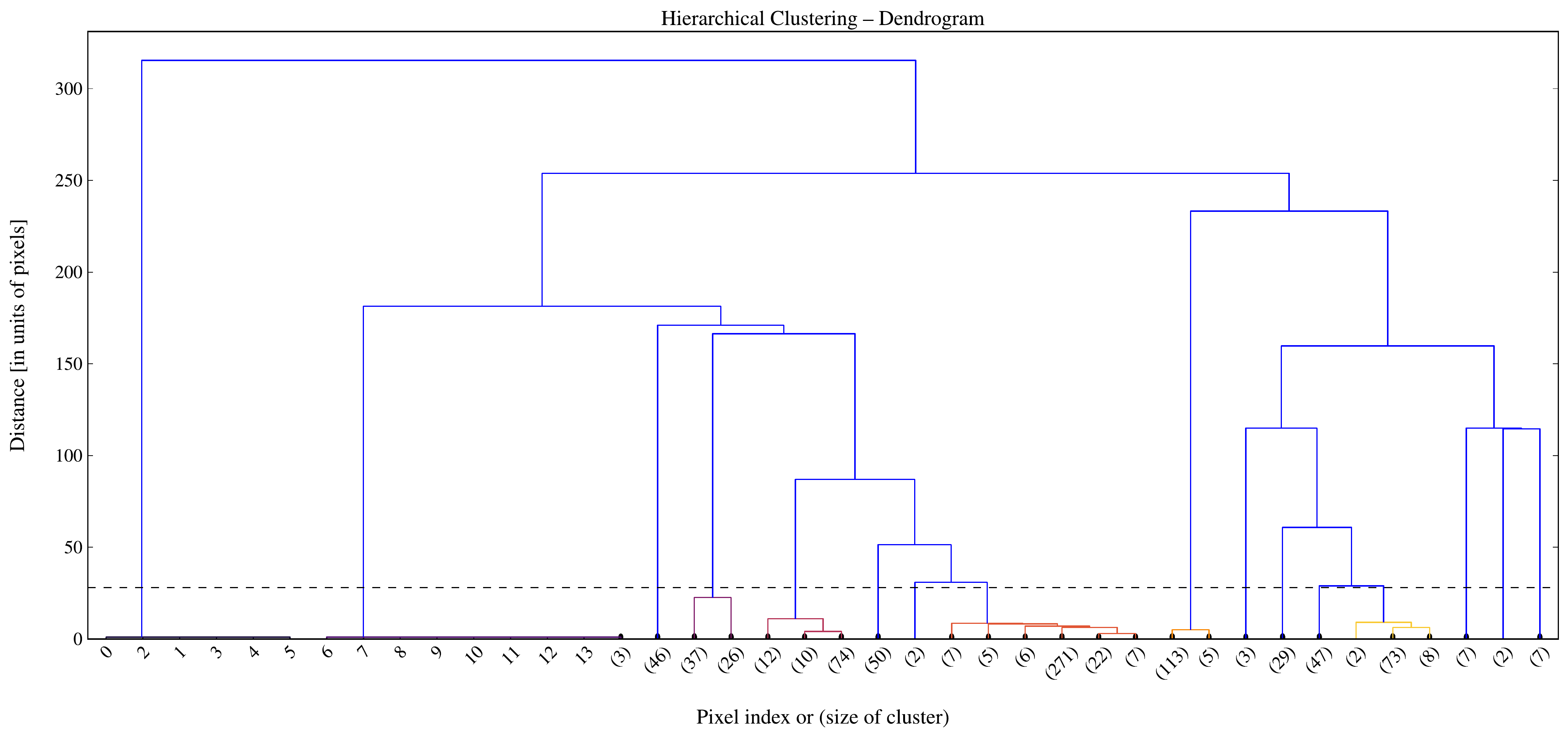}}
	\caption{Example output from the hierarchical clustering algorithm, showing transient candidates identified in the source extraction pipeline from the difference image corresponding to the 13\,s snapshot shown in (a). All pixels identified above a $5\sigma$ threshold in the region of the difference image shown in (b) are hierarchically clustered into individual sources, as shown visually in the dendrogram plot (c). The dashed line denotes the cutoff distance threshold used, above which all connecting nodes are disregarded and the clusters formed below this threshold represent the number of distinct sources identified in the difference image. In this example, there are 16 sources identified in the difference image, each of which is marked in (b) with a red ellipse.}
	\label{fig:cluster}
\end{center}
\end{figure*}

Sources that are present in each subtracted image are identified through the source extraction pipeline, cross-matched with any subsequent detections of a source at the same position in later snapshots and other timescales, and added to a candidate transient list. The source extraction algorithm is custom-built for the OVRO--LWA transient pipeline, and utilizes a hierarchical clustering algorithm (using \texttt{SciPy}'s \texttt{scipy.cluster} clustering package). This custom algorithm can accurately identify individual sources from the frequently noncontiguous set of pixels above the selected noise threshold that comprise a given source and its sidelobes in the non-deconvolved subtracted images. For this reason, the custom algorithm was selected over pre-existing source-finding algorithms that are frequently used with radio data. Candidate transient sources are identified through the following:
\begin{enumerate}
\item Each sequential subtracted image is divided up into 16 image regions, and all pixels above a $5\sigma$ local noise threshold in each region are identified and grouped into ``islands," using a hierarchical clustering algorithm with a distance-based linkage function (see Figure~\ref{fig:cluster}). There are on the order of 100 candidate transients identified in each sequential subtracted image above the $5\sigma$ local noise threshold.
\item The transient candidate detections in individual subtracted images are merged together across the full data set to create a catalog of transient candidates, which is cross-matched with the self-generated catalog of sources present within the 31\,hr data set (see Section~\ref{sourcecatalog}). The vast majority of sources detected in the sequential subtracted images are persistent sources that exhibit variability on 13\,s timescales, due to scintillation. Any remaining sources that are not co-located with any of the known cataloged sources are classified as potential transient events. This reduces the number of sources detected in a single difference image from $\sim$100 to a few. At this point in the pipeline, the threshold above which a source was kept in consideration as a candidate transient event was raised to $6.5\sigma$. This reduced the expected number of false-positive sources due to Gaussian noise fluctuations to $<$1 across the full data set, and maintained a manageable number of candidate events. Table~\ref{tab:noise} gives the approximate flux density levels to which this 6.5$\sigma$ threshold corresponds for all timescales.
\item Candidate sources that are detected more than 60 times across the full data set are classified as scintillating sources that were not captured by the source catalog, either because they were below the 4.5$\sigma$ detection threshold that was used to generate the catalog or they are positionally offset by more than 30$'$ from the cataloged source position due to ionospheric refraction, and are therefore removed from the candidates list. The 30$'$ offset was determined empirically from the data by measuring position offsets of point sources during periods of ionospheric activity and determining a typical refractive shift. In either of the above cases, the presence of a source is verified against a deeper reference catalog, including a deeper OVRO--LWA map~\citep{Eastwood+2018} and the VLA Sky Survey redux~\citep[VLSSr;][]{Lane+2014}.
\item The remaining candidate transient sources are compiled into a list, along with corresponding metadata, including S/N in the detection images, approximate source size, coordinates, azimuth and elevation at the time of detection, number of times a source was detected at this position, the spectrum of the source at 24~kHz resolution, and an automatically generated classification label that identifies the source type based on fitted source size, spectral features, and position in the beam (see Section~\ref{classification}). Table~\ref{tab:cands} shows the number of candidate transients remaining at this step in the pipeline, for each of the timescales searched.
\end{enumerate}

\begin{deluxetable}{ccc}[htb!]
	\tabletypesize{\scriptsize}
	\tablecolumns{3}
	\tablecaption{Number of Transient Candidates Remaining After Major Steps in the Transient Detection Pipeline. \label{tab:cands}}
	\tablehead{
		(1)					& (2)										& (3)		\\
		\colhead{Timescale}		& \colhead{$N_\text{candidates}$ After First Cut}		& \colhead{$N_\text{candidates}$ After Auto-classification}		\\
			}
	\startdata
		13\,s					& 4520		& 229		\\	
		39\,s					& 3761		& 152		\\	
		2\,minutes				& 3370		& 133		\\	
		6\,minutes				& 4669		& 598		\\	
		sidereal				& 1700		& 109		\\	
	\enddata
\end{deluxetable}

\subsection{Source Catalog}\label{sourcecatalog}
Due to the effects of the ionosphere (refraction and scintillation) and the solar wind~\citep[interplanetary scintillation; see, e.g.,][]{Kaplan+2015}, significant source variability is present on timescales of sub-integration and greater for the vast majority of sources in the OVRO--LWA FOV. In order to deal with the hundreds of (extrinsically) variable sources that are present in each sequentially subtracted image, an OVRO--LWA source catalog was generated (from the 31\,hr data set) in order to provide a self-consistent catalog of sources that can be cross-matched with sources detected in the subtracted images, thereby eliminating thousands of spurious detections due to extrinsic variability of persistent sources from consideration as transient sources. We note that this method of eliminating candidates that can be matched to a source in the catalog precludes detection of not only extrinsically but intrinsically varying sources. While intrinsic variability on timescales as short as tens of seconds from persistent sources in the catalog is not expected (e.g., incoherent synchrotron emission from active galactic nuclei in which light travel time arguments preclude the possibility of variability on such short timescales~\citealt{Pietka+2015}), probing intrinsic variability is of interest on the longest timescales (approximately a month) to which the search is sensitive. However, variability studies are beyond the scope of this current work.

The source catalog is constructed from snapshot images across the full 31\,hr data set that are separated by approximately 5\,hr, down to a maximum limiting elevation of 10$^\circ$. These snapshot images are median filtered in image space using a kernel size of 21 pixels, then subtracted from the nonfiltered image, in order to remove any large-scale diffuse galactic emission and improve the detection and fitting accuracy of point sources (above a 4.5$\sigma$ threshold) that can be recovered from the images and added to the source catalog. The source extraction algorithm used to generate the source catalog is the same algorithm used in the transient detection pipeline (see Section~\ref{sourceextraction}). The snapshot images used to construct the source catalog are filtered in image space rather than through the tapering of visibility weights because the image-space filter and subtraction remove extended emission without the ``ringing" introduced by the hard tapering of short-spacing visibilities that is necessary to fully remove diffuse emission.

From the 31\,hr data set, we assembled a catalog of 4500 sources that was used as a reference in the transient detection pipeline described below.

\subsection{Classification and Visual Inspection}\label{classification}
\begin{figure*}[htb!]
\begin{center}
\begin{tabular}{cc}
	\multirow{+22}[2]{*}{\subfigure[]{\includegraphics[width=0.45\textwidth]{f4a.pdf}}}	 & \\
	&	\subfigure[]{\includegraphics[width=0.5\textwidth]{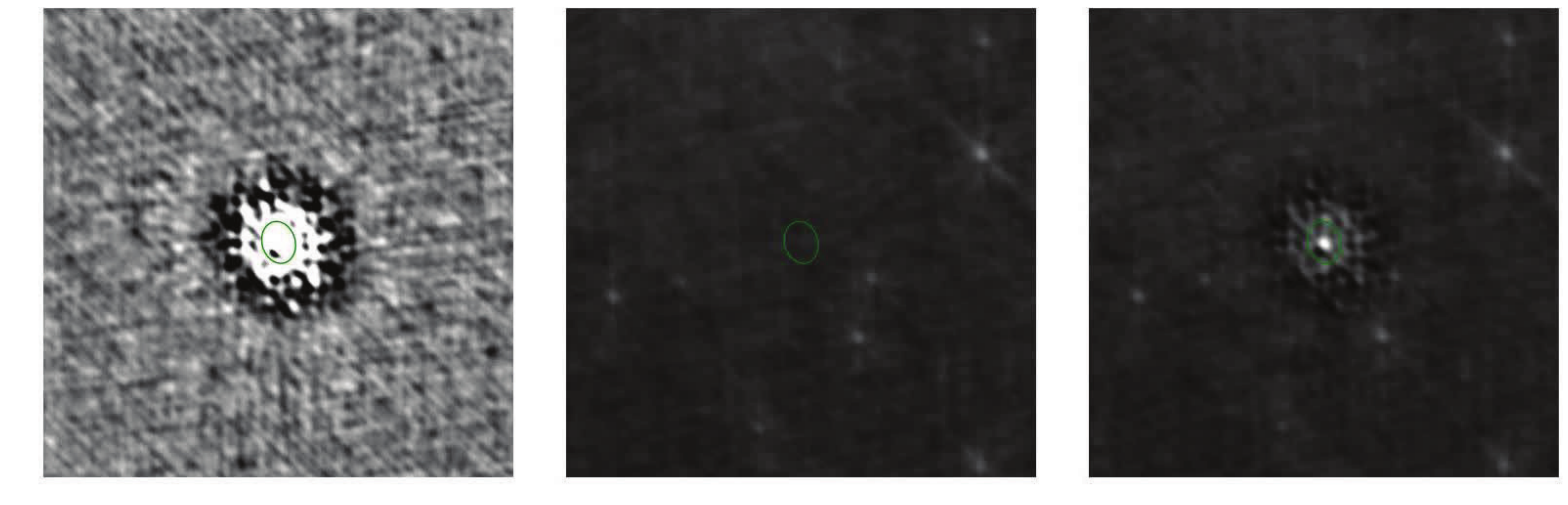}} \\
	&	\subfigure[]{\includegraphics[width=0.5\textwidth]{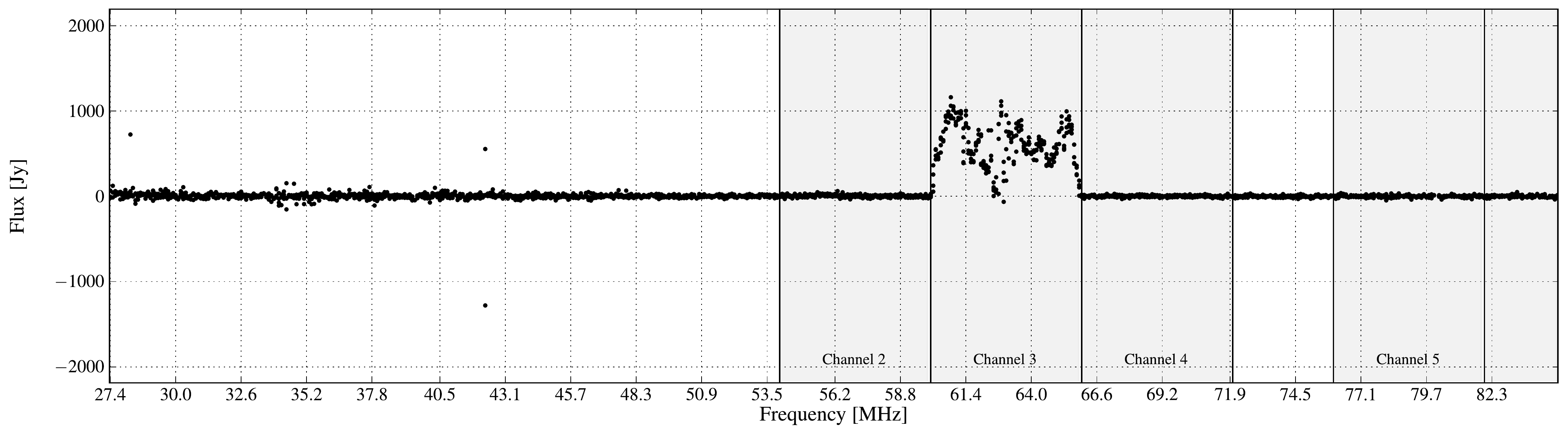}\label{fig:transientframespectrum}} \\
	&	\subfigure[]{\includegraphics[width=0.5\textwidth]{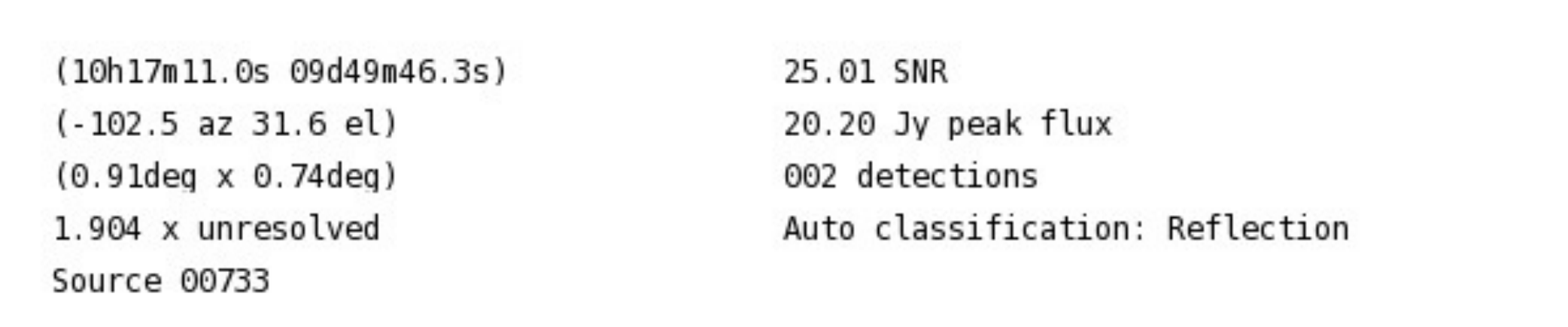}} \\
\end{tabular}
	\caption{An example output image from the transient pipeline. Panel (a) shows the subtracted image in which the candidate transient source is detected. The color scale on the image is from $-4$ to 4\,Jy. Black circles are covering sources that were detected in the source extraction pipeline but are present in the OVRO--LWA source catalog. Transient candidates without a catalog counterpart are labeled with their transient pipeline source ID. Image cutouts at the location of the transient candidate show the subtracted image in which it was detected, as well as the integrations that formed the subtracted image (b). The source difference spectrum (c) provides important diagnostic information, particularly in the case of meteor reflections, which are by far the most dominant nonastrophysical transient sources in the data set, occurring at a rate of approximately $0.1\,\text{s}^{-1}$. The vertical lines in the spectrum show the digital TV channel broadcast bands which are observable from the OVRO--LWA. In this example, the source spectrum is showing reflected Channel 3 TV broadcasts, which span 60--66\,MHz. The pipeline also outputs metadata (d) on the candidate transient source, including R.A.~and decl., azimuth and elevation, an approximate Gaussian fit to the source, the source ID, the S/N with which the source was detected, the peak flux (based on the approximate Gaussian fit), the number of times this source was detected, and an automatically generated classifier label.}
	\label{fig:transientframe}
\end{center}
\end{figure*}

Human visual inspection was necessary at this point in the pipeline, both to evaluate the success with which the source extraction algorithm recovered sources above the $6.5\sigma$ threshold in the subtracted images, and to ensure the accuracy of the automatically generated source classifiers. For each combination of timescale and data set, the transient pipeline outputs an image containing relevant diagnostic information for all remaining candidate transient events, which includes image cutouts, spectra, and relevant metadata. Figure~\ref{fig:transientframe} shows an example output frame from the transient pipeline---one such frame is generated for every transient candidate, in every integration in which said transient is detected. Because the source extraction algorithm selects both for positive and negative sources, each candidate event should be detected at least twice, as it will be present in two subtraction images---first as a positive source and then as a negative source. The vast majority of all candidate events fall into one of the following categories---RFI reflection events likely associated with meteor ionization trails, airplanes passing above the array (showing either reflected or self-generated RFI), broadband RFI on or within a few degrees of the horizon associated with power lines, Cas\,A and Cyg\,A sidelobes, and scintillating sources.

\subsection{Follow-up of interesting events}
Each candidate transient source is automatically given an initial classification based on the difference spectrum, fitted source size, and position. It is then visually inspected to ensure that classifiers are being correctly applied and that no transient source was missed. Eventually, the verification of sources by manual human inspection will be phased out of the pipeline and replaced with a more efficient and robust method of identifying and classifying sources using a machine learning algorithm. However, for the current data set, transient detection was automated to the point of candidate detection, which was sufficient for compiling a list of candidates and their associated metadata that could then be more robustly inspected and identified by human eyes. For those candidates that were not clearly associated with the types of events described above, and which did not have a counterpart in the VLSSr catalog at 74\,MHz, the TGSS~\citep{Intema+2017} catalog at 150\,MHz, or in the OVRO--LWA catalog generated from the (more sensitive) m-mode sky maps at 74\,MHz~\citep{Eastwood+2018}, the following steps were taken to further characterize the candidate and verify it as a transient event:
\begin{enumerate}
\item To better characterize the position, extent, and spectral behavior of each source, candidates are reimaged at phase center. This is done across the full 58\,MHz bandwidth, across multiple 2.6\,MHz subbands, and in the frequency range over which a source shows emission in the difference spectrum.
\item Deconvolution is performed with a range of visibility weightings between natural and uniform, in order to investigate the presence of flux on different spatial scales.
\item The source image is fit in order to confirm whether the source is unresolved, as expected for an astrophysical transient. If resolved, we investigate, typically confirming an atmospheric/meteoric reflection event.
The source visibilities are fit in order to robustly determine whether the source is unresolved---or, if it is resolved and in the near-field of the array, to determine the source height (in which case it is most likely an atmospheric / meteor reflection event).
\end{enumerate}

\section{Results}\label{results}
Across the entire 31\,hr data set, and for all timescales probed by the survey, a total of 36,232 images were searched in the transient pipeline, with each image covering approximately 17,000 deg$^2$ and nearly $5\times10^4$ independent beams. After eliminating all candidates associated with cataloged sources and repeat detections of sources at different timescales, 12,112 transient candidates were identified by the pipeline. Of these, 5828 events were associated with meteor reflections, 2011 events were associated with RFI reflected from and generated by airplanes, 1430 events were associated with horizon RFI or bright source sidelobes, and 1622 events were spurious detections of sidelobes or extended emission. None of the remaining 1221 events, which were not automatically classified by the pipeline into one of the previous categories, were identified as astrophysical transients following human inspection---the events included scintillation of known sources, as well as detections associated with the Sun and a number of Jovian bursts.

\subsection{Meteor Reflection Events}\label{meteor}
The most prevalent (nonastrophysical) transient source in our data set is associated with the reflection of RFI from patches of high ionization in the atmosphere, likely caused by meteors. Meteors passing through the atmosphere experience the ablation of material off their surface---this stripped material collisionally ionizes the surrounding air particles, creating a patch of ionized material with a plasma frequency $>100$\,MHz, and which is therefore able to reflect terrestrial RFI~\citep{Millman+1948, Greenhow1952}. These are easily distinguished as digital (and a likely a few remaining low-power analog) TV broadcasts by their spectral signature, with all emission fully contained within one or more of the broadcast bands designated by the Federal Communications Commission: Channel 2 (54--60\,MHz); Channel 3 (60--66\,MHz); Channel 4 (66--72\,MHz); Channel 5 (76--82\,MHz); Channel 6 (82--88\,MHz) (see~\citealt{Crane2008}). Figure~\ref{fig:transientframespectrum} shows a typical reflection spectrum associated with a meteor. While reflection events are consistently spectrally confined within one or more of the 6\,MHz wide digital broadcast bands, within those bands they can exhibit a wide range of spectral features, as well as a wide range in characteristic flux densities. They are typically less than 13\,s in duration, but particularly bright events can last as long as $\sim30$\,s. Multiple reflection sources that are adjacent in position and likely associated with the same meteor trail can have spectra that show emission in different broadcast bands.

The automatic classification of transient candidates associated with meteor reflections is determined by the spectrum and approximate fit to the size of the source. For every integration in which the source is detected, the spectrum is convolved with a 6\,MHz wide top-hat function, and the location of the maximum of this filtered spectrum is identified. If the maximum falls within one of the known broadcast bands, is above a manually set detection threshold, and the fit to the source in the non-deconvolved image is resolved (in the near-field regime of the array) and circular to within a factor of 2, then the source is automatically classified as a meteor reflection (the latter specification is to distinguish it from airplane reflections---see Section~\ref{airplane} below).

We note that none of the meteor-associated emissions we observed appeared to be intrinsic in nature, i.e. akin to the fireball radio afterglows observed by \citealt{Obenberger+2015b}. All meteor events observed by the OVRO--LWA in this data set were clearly reflection events from ionized trails. However, this is not inconsistent with the rate of fireball radio emission predicted by \citealt{Obenberger+2016}; for a flux density lower limit of 540\,Jy at 38\,MHz, they predict a detection rate of approximately 40 meteor afterglows year$^{-1}$.

\subsection{Airplanes}\label{airplane}
\begin{figure*}[htb!]
\begin{center}
\begin{tabular}{cc}
	\multirow{+22}[2]{*}{\subfigure[]{\includegraphics[width=0.45\textwidth]{f5a.pdf}}}	 & \\
	&	\subfigure[]{\includegraphics[width=0.5\textwidth]{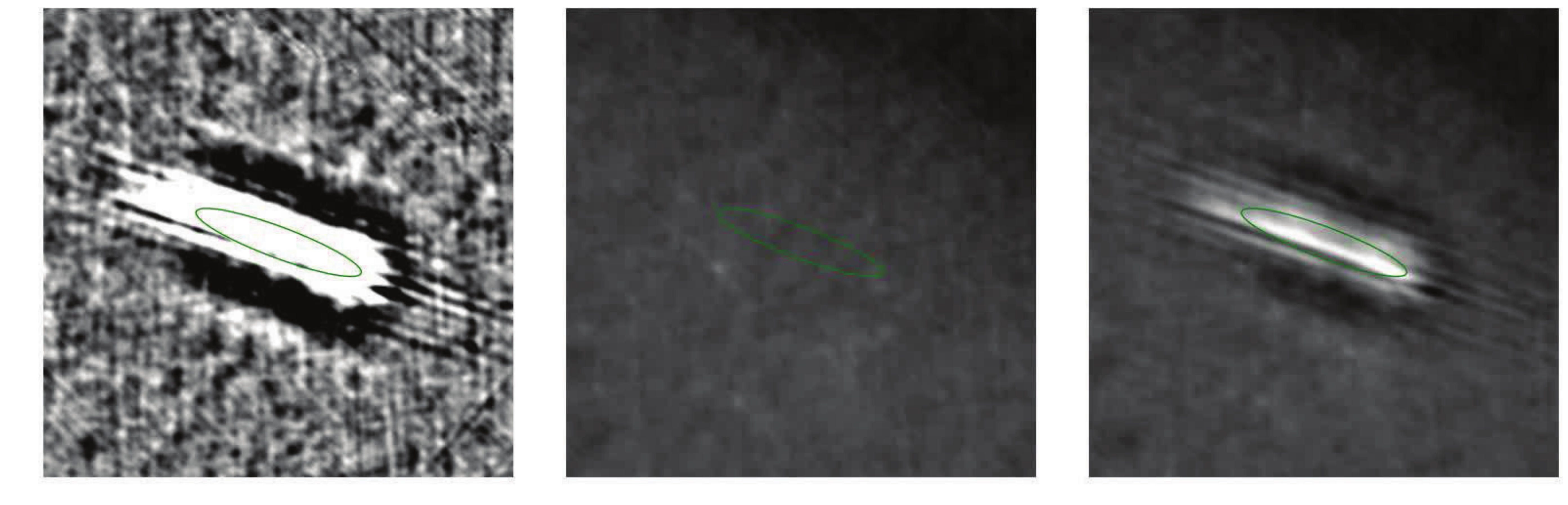}} \\
	&	\subfigure[]{\includegraphics[width=0.5\textwidth]{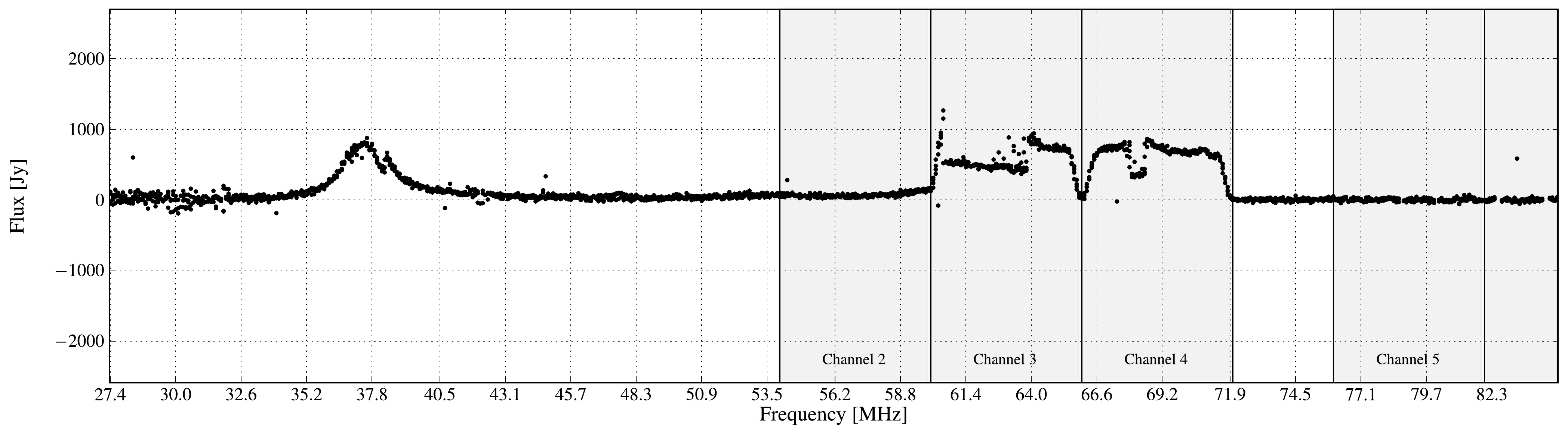}} \\
	&	\subfigure[]{\includegraphics[width=0.5\textwidth]{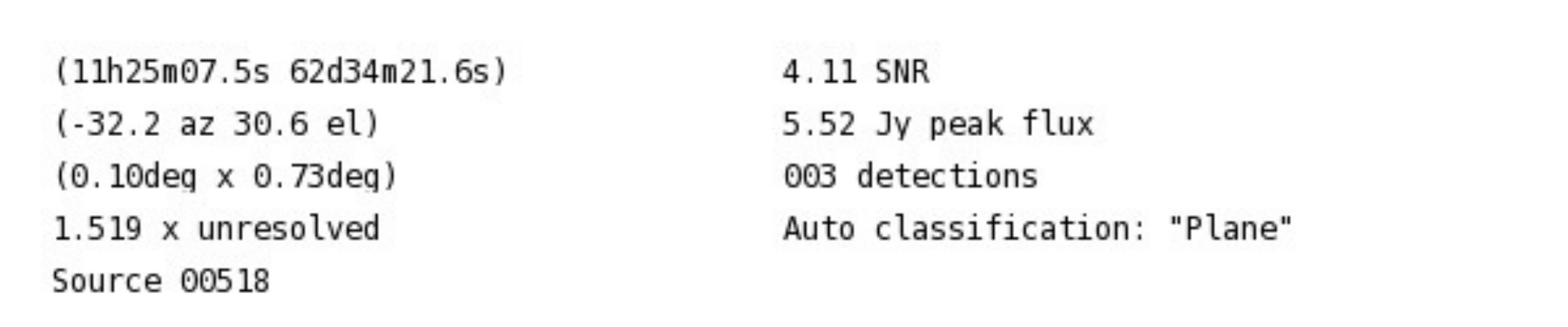}} \\
\end{tabular}
	\caption{Detection of an airplane in the transient pipeline. Airplanes are a persistent contaminant in the transient pipeline, but are easily identified by their temporal, spectral, and image features. This plane was detected to the northwest of the array, in the approximate direction of Bishop Airport. In this example, the plane is detected through both reflected RFI (digital broadcast bands in Channels 3 and 4) and RFI that is generated by the plane itself (the peak at approximately 37.5\,MHz).}
	\label{fig:airplane}
\end{center}
\end{figure*}

Airplanes are another prevalent terrestrial transient source. Like meteor trail reflections, they are most frequently detected through their reflection of digital TV broadcasts, although many do also exhibit spectral features outside the known digital TV broadcast bands that are likely intrinsic (see Figure~\ref{fig:airplane}). While airplanes are a major contaminant in the transient pipeline, they are easily identified and flagged by their spectral features, their elongated shape, and their movement across the FOV.

\subsection{Horizon RFI and Bright Sources}\label{horizon}
Sources of broadband RFI, which are distinct from the reflection of television broadcast bands, are confined to the horizon in discrete directions, usually along the line of sight to power transmission lines. Although the RFI is restricted to the horizon, it is still frequently detected by the transient pipeline at elevation angles as high as $\sim20^\circ$. However, because the RFI is localized, both in position and in time, it is classified in the transient pipeline by hierarchical clustering of sources located at a specific set of azimuths and low elevations, weighted by $1/\Delta t$ where $\Delta t$ is the time between separate detections.

A similar clustering scheme is used to flag detections in the transient pipeline associated with bright source sidelobes (e.g., Taurus\,A, Virgo\,A, Perseus\,A and B, etc.---typically sources with flux densities greater than approximately a few hundred Jy), as well as Cas\,A and Cyg\,A residuals.

\subsection{Scintillating Sources}\label{scintillation}
Another significant source of false positive transient candidate sources detected in the pipeline are quiescent sources that are varying due to scintillation. These include discrete sources that were previously below the detection limit of the source catalog, which scintillated above the detection threshold in one or more difference images, as well as sources that were detected and included in the source catalog but exhibited a refractive offset significant enough to not have been correctly identified by the transient pipeline as a known source (i.e. $>20'$ positional offset). In the latter case, all sources detected more than 60 times across the data set at a given timescale were automatically classified as such, and later manually verified as variable sources (due to scintillation effects) rather than true transient events. In the former case, the scintillating nature of the source as opposed to a transient is verified by fitting for the source position and flux across the full data set (and accepting a lower threshold than the $4.5\sigma$ cutoff that was used for generating the source catalog), as well as cross-matching the location of the source with the m-mode analysis maps generated from the same OVRO--LWA data set, and the VLSSr and TGSS catalogs.

A small subset of sources ($\sim10$) detected by the pipeline exhibited large increases in flux at frequencies that roughly correspond to the critical frequency of the ionosphere, at the transition between strong and weak scattering and where the magnitude of scintillation is greatest. The sources characterized by this type of scintillation behavior show increases in flux as large as 100s of Jy at frequencies between 30 and 40\,MHz, and evolving on 13\,s timescales and lasting for as long as a few minutes. The majority of these events occurred at low elevation ($<20^\circ$), but a few were detected at high elevation, usually during periods of increased scintillation due to ionospheric activity. The critical frequency occurs when the Fresnel and diffraction scales are approximately equal (see \citealt{Narayan1992}). The former scales with wavelength, $\lambda$, and distance between the observer and scattering screen, $D$, as
\begin{equation}
r_F = \sqrt{\frac{\lambda D}{2\pi}},
\end{equation}
\noindent and the latter as
\begin{equation}
r_{\rm{diff}} \propto \lambda^{\frac{6}{5}} D^{-\frac{3}{5}}
\end{equation}
\noindent under the assumption that the ionosphere is characterized by Kolmogorov turbulence. The diffraction scale of the ionosphere is approximately $1\,\rm{km}$ for a wavelength $\lambda \approx 300\,\rm{cm}$ and scattering screen distance of $D \approx 300\,\rm{km}$. The diffraction and Fresnel scales in the ionosphere are approximately equal at a frequency of $\sim 50\,\rm{MHz}$, which is roughly consistent with the frequencies at which we observe these events.

\section{Discussion}\label{discussion}
Our transient search across 31\,hr of OVRO--LWA data, at all five timescales probed, ranging from 13\,s up to sidereal day subtractions, resulted in no radio transients that could be identified conclusively as astrophysical in nature. We can use these nondetections to place upper limits on the instantaneous transient surface density for each timescale searched and at the flux density limits set by each timescale's sensitivity. The instantaneous transient surface density is the number of transient sources per square degree that would appear in a snapshot of the sky at a flux density greater than or equal to the flux density at which the surface density is reported. This is distinct from a transient rate, because the surface density is instantaneous and not reported per unit time. However, there is a timescale at which a transient surface density limit placed by a survey is applicable, and it is set by the cadence of the survey. We report the 95\% confidence level upper limit for the transient surface density, under the assumption that the rate of transients follows a Poisson distribution:

\begin{equation}\label{eq:poisson}
P(n) = \frac{\lambda^n}{n!} e^{-\lambda},
\end{equation}

\noindent where $\lambda=\rho\Omega_\text{tot}$ is the expected number of transients on the sky at any instant in time, $\rho$ is the instantaneous transient surface density, and $\Omega_\text{tot}=(N_\text{ind.\,epochs} - 1) \times \Omega_\text{FOV}$ is the total surface area probed by the survey. For the OVRO--LWA, the FOV that is searched in each epoch is 17,045$\,\text{deg}^{-2}$. Here, $N_\text{ind.\,epochs}$ is the number of independent epochs searched (i.e., epochs separated by at least the transient timescale probed), which for our survey is smaller than the total number of epochs searched by a factor $t_\text{int}/t_\text{probed}$. This is due to the fact that our search on longer timescales was conducted on difference images that were generated from two integrations separated by the timescale being probed, and which were shifted by single integration timescales to generate subsequent difference images. The result is oversampling in the time domain of transients on timescales longer than the integration time. This is advantageous for maintaining sensitivity to potential transient events by providing a more matched temporal filter, but it also results in many epochs that are not independent samples of the transient sky (because they are separated in time by less than the timescale being probed).

\begin{figure*}[htb!]
\begin{center}
	\includegraphics[width=1\textwidth]{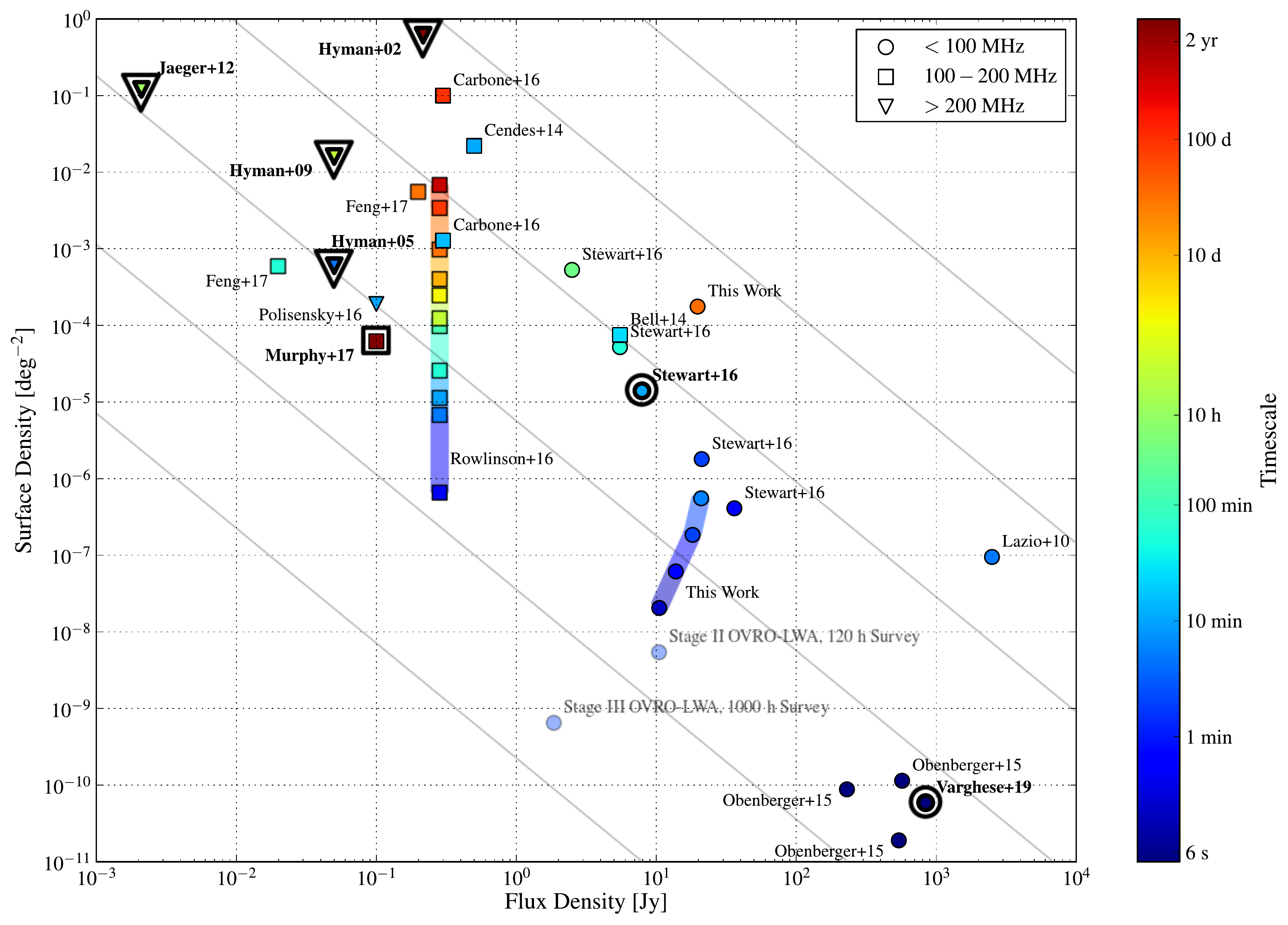}
	\caption{Radio transient phase space plot, showing the transient surface density as a function of flux density limits and timescales probed, for this and previous blind transient surveys (see Table~\ref{tab:previoussurveys} for an overview of all surveys included in this plot). The surface density limits for each survey are colored according to the timescale probed, covering timescales as short as 5\,s \citep{Obenberger+2015} to as long as 3\,yr \citep{Murphy+2017}. This is a critical parameter, as surveys providing surface density limits at comparable sensitivities may be probing very different timescales and therefore very different regions of phase space and potential transient populations. The same is true of frequency. Surveys conducted below 100\,MHz are marked with a circle, those between 100 and 200\,MHz are marked with a square, and those between 200 and 350\,MHz with a triangle. All points on the plot denote upper limits with the exception of the seven transient detections that are marked in bold. The solid gray lines denote hypothetical transient populations under the generic assumption that the population is a standard candle in a Euclidean universe, i.e., $N(>S) \propto S^{-\gamma}$, where $N$ is the number density of sources, $S$ is the flux density, and $\gamma=3/2$. The limits (or detections) placed by transient surveys are necessarily a combination of sensitivity and total area surveyed, and it is often the case that tradeoffs must be made to improve one of these factors over the other. The optimal combination of these two parameters (and therefore how deep or wide a survey probes) depends on the goals of the survey or the source population(s) it is targeting. Also shown in this figure are the limits we expect to achieve with the OVRO--LWA in future surveys, first with the 120\,hr transient survey with the stage II OVRO--LWA, and finally with the completed stage III OVRO--LWA utilizing 1000\,hr of data capable of achieving 150\,mJy snapshot sensitivity.}
	\label{fig:phasespace}
\end{center}
\end{figure*}

From our survey nondetection ($n=0$), we can place upper limits on the instantaneous transient surface density at a 95 percent confidence level ($P(0) = 0.05$), for all timescales that were searched, using Equation~\ref{eq:poisson}. Table~\ref{tab:limits} shows the instantaneous surface density limits placed for all timescales probed by the survey. The surface density limits placed, in the context of previous transient searches, is shown in Figure~\ref{fig:phasespace}. We have placed the most constraining limits on the transient surface density at low frequencies at $\sim$few Jy-level sensitivities. In addition, our upper limits are consistent with those placed by \citealt{Obenberger+2015} at comparable frequencies and flux densities of a few hundred Jy, as well as with those placed by \citealt{Rowlinson+2016} at 182\,MHz, assuming a flat spectral index and a standard candle population of transients in a Euclidean universe.

\begin{deluxetable}{cccc}[htb!]
	\tabletypesize{\scriptsize}
	\tablecolumns{4}
	\tablecaption{Transient Surface Density Limits Placed for All Six Timescales Probed by the OVRO--LWA Transient Survey \label{tab:limits}}
	\tablehead{
		(1)					& (2)							& (3)							& (4)					\\
		\colhead{Timescale}		& \colhead{$N_\text{ind.~epochs}$}	& \colhead{Sensitivity (6.5$\sigma$)}	& \colhead{$\rho$}		\\
							& 							& \colhead{(Jy)}				& \colhead{(deg$^{-2}$)}	\\
			}
	\startdata
		13\,s					& 8586						& 10.5						& $2.50\times10^{-8}	$	\\
		39\,s					& 2862						& 13.8						& $6.14\times10^{-8}$	\\
		2\,minutes				& 954						& 18.1						& $1.84\times10^{-7}$	\\
		6\,minutes				& 318						& 20.8						& $5.53\times10^{-7}$	\\
		sidereal (1\,day)			& 1							& 19.3						& $1.76\times10^{-4}$	\\
		sidereal (35\,days)			& 1							& 19.7						& $1.76\times10^{-4}$	\\
	\enddata
	\tablecomments{Because the entire OVRO--LWA FOV down to an elevation angle of $10^\circ$ is searched as part of the transient pipeline, our single snapshot FOV is approximately 1.65$\pi$\,sr, or 17,045$\,\text{deg}^2$.}
\end{deluxetable}

We can also place limits on the transient surface density as a function of flux density, $\rho(S)$, rather than reporting the surface density at a single flux density value approximated as an average of our survey sensitivity over the primary beam. Because the transient search is conducted over nearly the entire primary beam FOV, each epoch of our survey has nonuniform sensitivity, with the best and worst sensitivities achieved at zenith and $10^\circ$ elevation angle, respectively. However, for a transient population described by a luminosity function with power law $\gamma$, varying sensitivity corresponds to probing different parts of the luminosity distribution, and therefore the number of transients we can expect to detect within the volume probed by our survey is a function of sensitivity and the corresponding fraction of the sky covered by each sensitivity (see, e.g., the methods presented in~\citealt{Carbone+2016}). Using Equation~\ref{eq:poisson}, we can write

\begin{equation}\label{eq:poissonphi}
P(0) = e^{-\rho(S(\phi))~\Omega_\text{tot}(\phi)},
\end{equation}

\noindent where $\rho(S(\phi)) = \rho_o \left( \frac{S(\phi)}{S_o} \right)^{-\gamma}$ and $\Omega_\text{tot}(\phi) = (N_\text{ind.\,epochs} - 1) \times \Omega_\text{FOV}(\phi)$. The flux density to which we are sensitive is a function of zenith angle $\phi$, and determined by the primary beam gain pattern, which we approximate as $\cos^{1.6}(\phi)$. We can therefore write Equation~\ref{eq:poissonphi} as 

\begin{equation}
P(0) = e^{-\int \rho_o \left( \frac{S_\text{zenith} / \cos^{1.6}\phi}{S_o} \right)^{-\gamma} \times~(N_\text{ind.\,epochs} - 1)~2\pi \sin\phi d\phi},
\end{equation}

\noindent where $S_\text{zenith} = 6.5\sigma_\text{zenith}$ is our transient survey detection limit at zenith. Under the assumption of $\gamma = 3/2$ for a standard candle population distributed uniformly in Euclidean space, the 95\% confidence level on the transient surface density as a function of flux density, $\rho_o(S_o)$, becomes

\begin{equation}
\rho_o(S_o) = -\frac{\ln{0.05}}{(N_\text{ind.\,epochs} - 1)~2\pi \int_{0^\circ}^{80^\circ} \cos^{1.6\gamma}\phi~\sin\phi~d\phi} ~\left( \frac{6.5\sigma_\text{zenith}}{S_o} \right)^\gamma,
\end{equation}

\begin{equation}
\rho_o(S_o) = - \frac{\ln{0.05}}{(N_\text{ind.\,epochs} - 1)~0.6\pi} ~\left( \frac{6.5\sigma_\text{zenith}}{S_o} \right)^{3/2} \text{sr}^{-1},
\end{equation}

\noindent where $\sigma_\text{zenith}$ for each timescale probed is given in column 2 of Table~\ref{tab:noise}.

\begin{figure*}[htb!]
\begin{center}
	\includegraphics[width=0.8\textwidth]{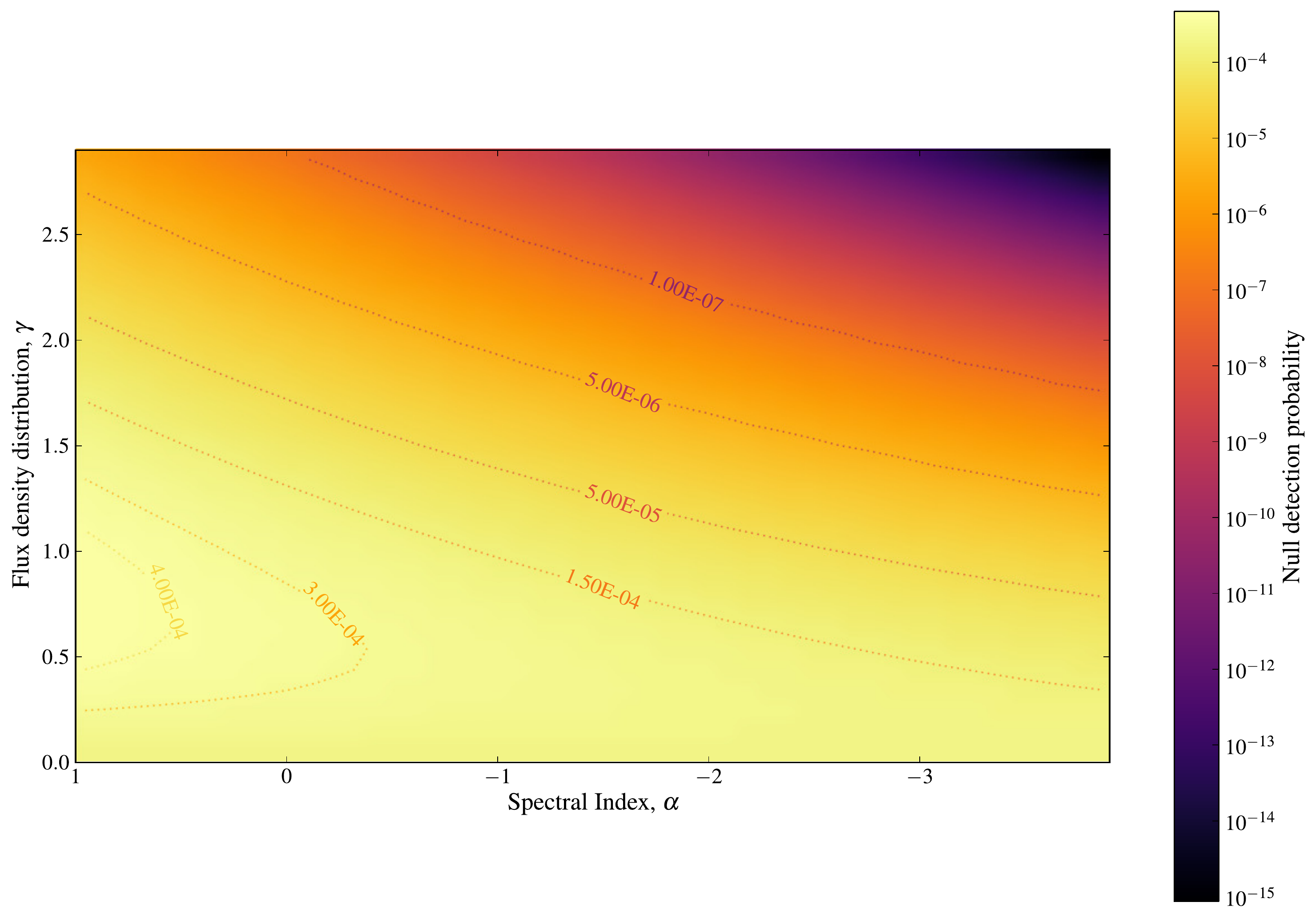}
	\caption{Probability of a null detection in the OVRO--LWA 31\,hr survey of the potential population indicated by the event detected by \citealt{Stewart+2016}, given their transient rate of $3.9\times10^{-4}\,\text{days}^{-1}\,\text{deg}^{-2}$, as a function of the power-law luminosity distribution $\gamma$ and source spectral index $\alpha$, using Equation~\ref{eq:poissonphi}.}
	\label{fig:nullprob}
\end{center}
\end{figure*}

\begin{figure*}[htb!]
\begin{center}
	\includegraphics[width=0.8\textwidth]{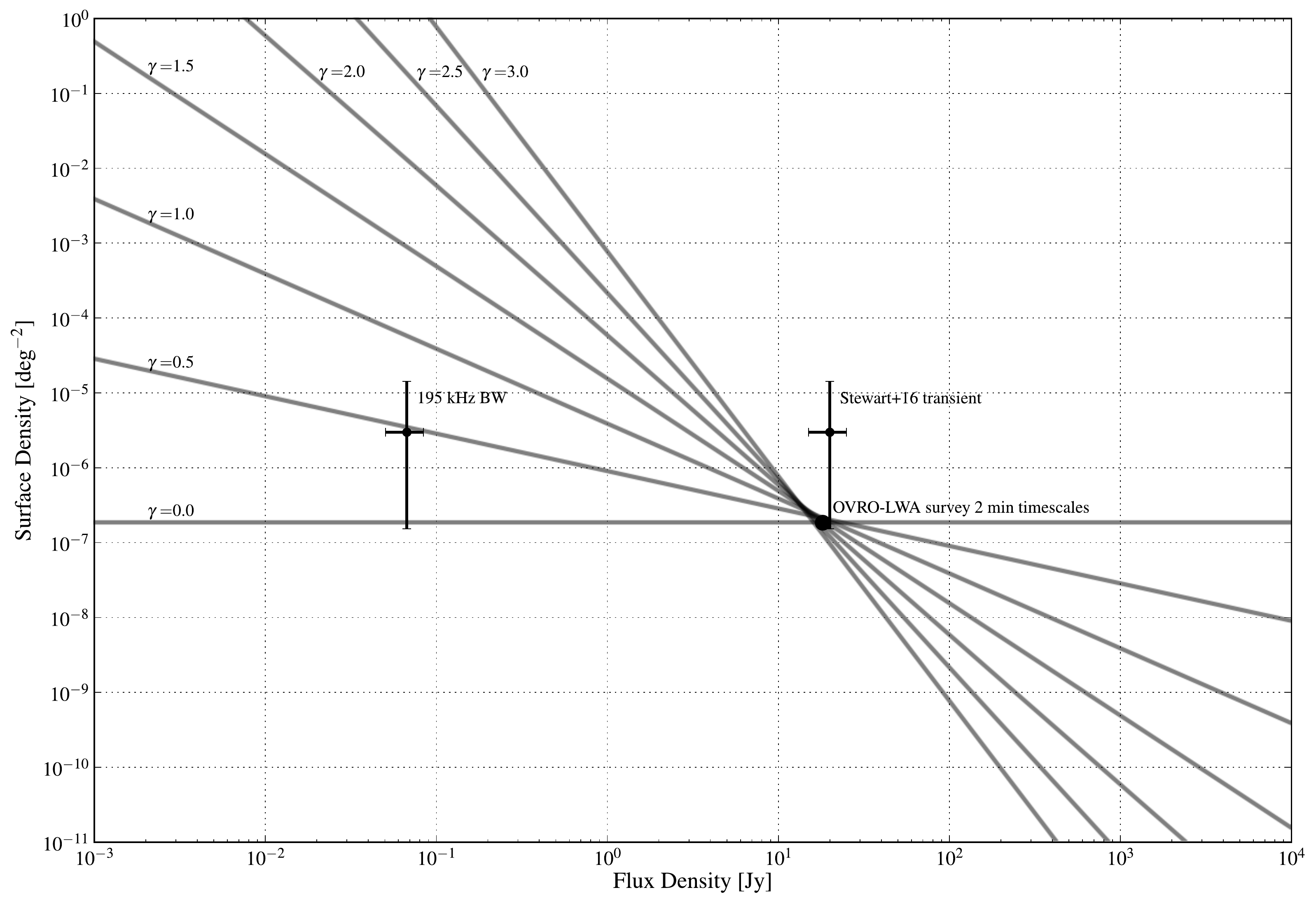}
	\caption{Surface density limits from the OVRO--LWA transient survey at two-minute timescales, for a range of power-law luminosity distribution $\gamma$. For reference, the surface density for the transient detected by \citealt{Stewart+2016}, at a reported flux density of between 15 and 25\,Jy, and at the flux density that same event would appear at in our survey for a maximum intrinsic bandwidth of 195\,kHz (the bandwidth of the survey by \citealt{Stewart+2016}). See also Figure 4 of~\citealt{Carbone+2016} for constraining transient surface density as a function of flux density for different power-law luminosity distributions.}
	\label{fig:stewartetaltransient}
\end{center}
\end{figure*}

With the exception of the five transient detections at $>150$\,MHz by \citealt{Murphy+2017}, \citealt{Jaeger+2012}, and \citealt{Hyman+2002, Hyman+2005, Hyman+2009}, and the detections by \citealt{Stewart+2016} and \citealt{Varghese+2019} of transients at 60 and 34\,MHz, respectively, all of the transient surveys shown in Figure~\ref{fig:phasespace} were only able to place upper limits on the transient rate at low frequencies. We note that, in the case of the three detections by \citealt{Hyman+2002, Hyman+2005, Hyman+2009}, all were found in targeted observations of the galactic center. Based on their brightness temperatures and the very low spatial volume covered by these surveys, these detections are most likely located in the galactic center and therefore extrapolation of the detection of these objects to an all-sky rate is not applicable. In the case of the detection by \citealt{Jaeger+2012}, the transient has a much longer timescale than what is probed in the bulk of our survey, and was also found at significantly higher frequencies, indicating a potential source population to which our survey would not be sensitive. The transient found by \citealt{Stewart+2016} was detected at the same frequency and timescale probed by the OVRO--LWA transient survey and at a comparable sensitivity, with a reported rate of $3.9^{+14.7}_{-3.7}\times10^{-4}\,\text{days}^{-1}\,\text{deg}^{-2}$, corresponding to a detection of $8.4^{+31.8}_{-8.0}$ such events in our survey. However, its detection by LOFAR at 60\,MHz with a 200\,kHz bandwidth cannot constrain the spectral index or intrinsic bandwidth of emission for this event. The probability of our null detection of this potential transient population is $1.9^{+644}_{-1.9}\times 10^{-3}$, which allows us to rule out a transient rate $> 1.4\times10^{-4}\,\text{day}^{-1}\,\text{deg}^{-2}$ with 95\% confidence. However, this is under the assumption of a flat spectral index and that the emission is broadband in nature. Figure~\ref{fig:nullprob} shows the probability of our null detection for a range of assumed spectral indices $\alpha$ and flux density distributions $\gamma$. If, however, the emission is of a narrow bandwidth relative to our observations, our sensitivity to this event decreases by a factor $\Delta\nu_\text{intrinsic}/\Delta\nu_\text{obs}$, where $\Delta\nu_\text{obs}=58\,\text{MHz}$ is our survey bandwidth. Figure~\ref{fig:stewartetaltransient} shows the transient surface density limits placed by our survey as a function of flux density, for a range of flux density distributions $\gamma$.

\section{Conclusion and Future Directions}\label{conclusion}
We have placed the most constraining limits on the transient surface density on timescales of 13\,s to a few minutes and at frequencies below 100\,MHz, with 31\,hr of data from the OVRO--LWA. No transients were discovered on timescales of 13\,s, 39\,s, 2\,minutes, or 6\,minutes, nor in the 7\,hr of data that had a corresponding overlap in local sidereal time (LST) and allowed us to search up to timescales as long as 1 to 35\,days. From our nondetections, we place an upper limit to the transient surface density of $2.5\times10^{-8}\,\text{deg}^{-2}$ at the shortest timescales probed at a flux density of 10.5\,Jy.

The nondetection of any sources akin to the transient event detected by \citealt{Stewart+2016} allows us rule out a rate $> 1.4\times10^{-4}\,\text{days}^{-1}\,\text{deg}^{-2}$ with 95\% confidence, under the assumption of a flat spectrum and wide bandwidth for this event. We further rule out a range of power-law luminosity distributions and emission spectral indices for the potential population detected by \citealt{Stewart+2016}, indicating that the event is likely narrow-band in nature. We note, however, the danger of conducting putative population rates with a transient detection of one -- the transient uncovered by \citealt{Stewart+2016} could indeed have been the first detection of a population, but a uniquely bright outlier from that population. The event rate drawn from that single detection could therefore be vastly overestimated.

The nondetection of the population implied by the transient detected by \citealt{Stewart+2016} can be used to inform the design of future planned surveys with the OVRO--LWA. The next transient survey will use additional data taken over a continuous 120\,hr of observations with the Stage II OVRO--LWA in 2018 March. The duration of this data set enables the transient search pipeline to operate entirely on difference images formed from sidereally matched integrations. This allows us to probe all timescales (between the integration time of 13\,s and the maximum separation of identical LSTs of five days) simultaneously, while avoiding the issues associated with sidelobe confusion and incomplete sky subtraction for diffuse emission away from zenith that cause an increase in noise on longer timescales. This transient survey will also feature sub-band searches in order to provide a filter in frequency space that is better matched to the transient detected by \citealt{Stewart+2016}, as well as blind transient injections to better constrain the transient surface density probed as a function of sensitivity, timescale, and frequency parameters.

Finally, the stage III OVRO--LWA, which is scheduled to begin construction in 2019, will provide orders of magnitude more sensitivity to transient events, with a stated goal of 1000\,hr for the first transient survey with the completed array, and 150\,mJy sensitivities. In addition, mapping of individual dipole beams (a stated technical goal for the final stage array) will eliminate the sidelobe confusion that necessitates peeling, and will enable us to integrate down in order to achieve better sensitivity on timescales longer than the 13\,s integration time.

\acknowledgements The authors would like to thank the anonymous referee for useful and constructive comments that helped us to improve the original text of this paper. This material is based in part upon work supported by the National Science Foundation under Grant AST-1654815 and AST-1212226. G.H.~acknowledges the support of the Alfred P.~Sloan Foundation and the Research Corporation for Science Advancement. The OVRO--LWA project was initiated through the kind donation of Deborah Castleman and Harold Rosen.

Part of this research was carried out at the Jet Propulsion Laboratory, California Institute of Technology, under a contract with the National Aeronautics and Space Administration, including partial funding through the President's and Director's Fund Program.

\software{CASA \citep{McMullin+2007},
			  TTCal calibration software package for the OVRO--LWA \citep{eastwood_michael_w_2016_1049160}, 
                           WSClean \citep{2014ascl.soft08023O},
                           Scipy \citep{Jones+2001}			}

\bibliography{references}

\end{document}